\documentclass[12pt, draftclsnofoot, onecolumn]{IEEEtran}
\usepackage{latexsym}
\usepackage{graphicx}
\usepackage{amsfonts,amssymb,amsmath}
\usepackage{hyperref}
\def\BibTeX{{\rm B\kern-.05em{\sc i\kern-.025em b}\kern-.08em T\kern-.1667em\lower.7ex\hbox{E}\kern-.125emX}}

\usepackage{graphicx}
\usepackage{amsmath,bbm,epsfig,amssymb,amsfonts, amstext, verbatim,amsopn,cite,subfigure,multirow,multicol,lipsum}
\usepackage{balance}
\usepackage{url}
\usepackage{amsfonts}
\usepackage{epsfig}
\usepackage{epstopdf}
\usepackage{setspace}
\usepackage{stmaryrd}
\usepackage{psfrag}	
\usepackage{multirow}
\usepackage{float}
\usepackage{cases}  
\usepackage[process=auto]{pstool}
\usepackage{etoolbox}
\usepackage{algorithm}
\usepackage{algorithmic}
\usepackage{hyperref}
\usepackage{wasysym}
\usepackage{pifont}
\usepackage[shortlabels]{enumitem}
\allowdisplaybreaks
\usepackage{breakurl}

\newtheorem{theo}{Theorem}
\newtheorem{lem}{Lemma}
\newtheorem{remk}{Remark}


\EndPreamble
\begin{document}
\newtoggle{OneColumn}

\toggletrue{OneColumn}

\title{Ergodic Sum Rate Analysis of  UAV-Based Relay Networks with Mixed RF-FSO Channels}

\author{Hedieh Ajam, Marzieh Najafi, Vahid Jamali,    and Robert Schober
}
\maketitle
\begin{abstract}
Unmanned aerial vehicle (UAV)-based communications is a promising new technology that can add a wide range of new capabilities to the current network infrastructure. Given the flexibility, cost-efficiency, and convenient use of UAVs, they can be deployed as temporary base stations (BSs) for on-demand situations like {BS} overloading or natural disasters. In this work, a UAV-based communication system with radio frequency (RF) access links to the mobile users (MUs) and a free-space optical (FSO) backhaul link to the ground station (GS) is considered. In particular,  the RF and FSO channels in this network depend on the UAV's positioning and (in)stability. The relative position of the UAV with respect  to the MUs impacts the likelihood of a line-of-sight (LOS) connection in the RF link and the instability of the hovering UAV affects the quality of the FSO channel. Thus, taking these effects into account, we analyze the end-to-end system performance  of networks employing UAVs as buffer-aided ({BA}) and non-buffer-aided (non-{BA}) relays in terms of the ergodic sum rate. Simulation results validate the accuracy of the proposed analytical derivations and reveal the benefits of buffering for compensation of the random fluctuations caused by the UAV's instability. Our simulations also show  that the ergodic sum rate of both BA and non-BA UAV-based relays can be enhanced considerably by optimizing the positioning of the UAV. We further study the impact of  the MU density and the weather conditions on the end-to-end system performance.
\end{abstract}

\section{Introduction}
\IEEEPARstart{F}{ifth} generation (5G) and beyond wireless communication networks are expected to overcome many of the shortcomings of the current infrastructure by offering higher data rates,  improving the quality of service (QoS) in  crowded areas, and reducing the blind spots of  current networks \cite{5G}.  Among other techniques, unmanned aerial vehicles (UAVs) have been introduced to  achieve the aforementioned goals. The unique characteristic of UAVs is their flexible positioning which together with their  cost efficiency and easy deployment makes them promising candidates for  a wide range of applications. For example, they may be used as relays  for coverage enhancement \cite{Drone_coverage},  as temporary base stations (BSs) for on-demand situations \cite{TempBS}, for adaptive fronthauling/backhauling \cite{FSO_Drone_UBC}, and for data acquisition for the Internet-of-Things (IoT) \cite{IoT}. Despite their expected benefits, the backhaul/fronthaul links needed to connect a UAV with a ground station (GS) constitute a  major challenge in UAV-based networks. The authors  of \cite{Wifi_backhaul} considered WiFi and satellite links for backhauling. However, for many applications,  UAVs have  to transfer huge amounts of data to the GS and  the backhaul links have to be able to cope with the UAV's mobility and the interference from other UAVs and the mobile users (MUs).  To address these issues, the authors of  \cite{FSO_Drone_UBC} and  \cite{FSO_Drone_Alouni} proposed free-space optical (FSO) systems   for the fronthaul/backhaul connections in UAV-based networks. FSO links offer high data rates (up to $10\ \mathrm{Gbps}$) by using the optical range of the frequency spectrum. Moreover, FSO systems are not suceptible to interference owing to  their narrow laser beams and are able to communicate over large distances (few kilometers) \cite{Survey_Khalighi}.

 Considering the above advantages,  the QoS and coverage of the current  infrastructure can be improved   by UAVs operating as  flying relays. Particularly, UAVs can be deployed as mobile relay nodes that forward the huge amounts of  data collected via RF access links from MUs to a GS via   FSO backhaul links.  The viability of   UAV-based communication systems has been demonstrated recently in  \cite{R2} and \cite{R3}, where end-to-end long term evolution (LTE) connectivity was provided by low altitude UAVs and balloons, respectively. Furthermore,   UAVs with FSO backhauling  were utilized in the  Aquila \cite{Aquila} and Loon \cite{Loon} projects for providing connectivity to the remote parts of the world. 
However, the performance of such  UAV-based relay networks has not been studied in detail and  is expected to be strongly dependent on the UAV's positioning and (in)stability. In particular, the location of the UAV with respect to (w.r.t.) the MUs and GS and its random vibrations in the hovering state affect the quality of the RF and FSO channels. Thus, the  impact of the UAV's positioning and instability on the following parameters necessitates  a careful study of  the performance of the end-to-end network:

\begin{enumerate}
\item \textit{MU distribution:} The {MU}s are randomly distributed and their data traffic  patterns may change over time. The  flexible positioning  of the UAV above the randomly distributed {MU}s can reduce  path loss and shadowing effects in the RF access links and boost the end-to-end system performance. 
\item  \textit{Line-of-sight ({LOS}) link:} The probability of maintaining a {LOS} path for the {RF} access links depends on  the elevation angle of the UAV w.r.t.  the {MU}s. When the UAV is at  higher altitudes, the {LOS} path between the UAV and a given {MU} is less likely to be blocked. Hence, depending on the position of the UAV, the distribution of the RF access channel coefficients can be either Rician in the presence of a LOS path  or Rayleigh  in the absence of a {LOS} path. Thus, the positioning  of the UAV determines  the LOS probability. 
\item \textit{Quality of the {FSO} link:}  In UAV-based {FSO} communications,  tracking errors and the instability of the hovering UAV degrade the intensity of the optical signal received at the photo detector (PD) of the GS. Furthermore, the distance between  the UAV and the GS affects the atmospheric loss in the FSO channel.
\end{enumerate}

The above factors have to be taken simultaneously into account for  the design of UAV-based networks. For instance, the position of the UAV w.r.t. the MUs affects both  the LOS probability of the access links and the atmospheric loss in the backhaul channel.  However, in previous works, only some of the above aspects were considered. For example, the authors of  \cite{Throughputmax} considered only the impact of the RF access links and assumed a perfect backhaul connection. They  determined the optimal placement of a stationary UAV and the optimal trajectory of a moving UAV in terms of the maximum throughput. Furthermore,  the performance of a cluster-based UAV network in terms of coverage probability  and energy efficiency was analyzed in \cite{TempBS}. The authors of   \cite{Place_Mozaffari}  investigated the  optimal positioning in a multi-UAV network with the objective to  minimize  the  total transmit power. In both \cite{TempBS} and \cite{Place_Mozaffari}, the backhaul link was assumed to be ideal. The authors of \cite{Rev:Drone_backhaul} considered the impact  of the positioning of the UAV on both the access and the backhaul links. However, in \cite{Rev:Drone_backhaul}, despite the potentially higher data rates of FSO links, an RF link was considered  for backhauling and the position of the UAV and the  RF bandwidth shared between  the access and backhaul links was optimized.  In fact, to the best of the authors' knowledge, the performance  of a UAV-based network employing FSO backhaul and  RF access links to connect  randomly distributed MUs to a GS  has not been studied  in the literature, yet. 

In this paper, we consider a UAV-based network where a hovering UAV acts as a decode-and-forward (DF) relay and connects   Poisson distributed MUs via RF access links  and   an FSO backhaul link  to a fixed GS. Because of the mutual orthogonality of the RF and FSO links, the relaying UAV can concurrently transmit and receive. Moreover,  depending on whether the data is delay sensitive or not, non-buffer-aided (non-BA) and buffer-aided (BA) relaying UAVs are considered, respectively  \cite{vahid_FSOrelay, vahid_TWC_partI, vahid_TWC_partII}. In particular,  BA relaying allows the UAV to transmit, receive, or simultaneously  transmit and receive depending on the channel conditions \cite{Nicola_FDBA}.  The performance of the  considered UAV-based relay network is analyzed in terms of the ergodic sum rate for both BA and non-BA relaying. We  validate the proposed analytical results with computer simulations.  The main contributions of this paper can be summarized as follows:
\begin{itemize}
	\item \textbf{End-to-end system model including both access and backhaul links:} We investigate the end-to-end performance of UAV-based relay systems and show that the performance is simultaneously dependent on both the access and  backhaul links. Thereby,  if one link degrades the end-to-end performance, the position of the UAV can be adjusted accordingly to enhance the performance. 
	\item \textbf{Impact of UAV positioning and (in)stability:}  The end-to-end ergodic sum rate of the system is  analyzed taking into account the impact of  the  positioning of the UAV w.r.t. the MUs and the GS and the random fluctuations of the position and  orientation of the UAV in the hovering state which in turn affect the quality of the access and backhaul links. We show that these characteristics of  UAVs have to be jointly considered for  performance analysis of UAV-based communication networks.
	\item \textbf{Impact of buffering on  performance:} We analyze the ergodic sum rate for both BA and  non-BA  UAV-based relay systems and  show that buffering can mitigate  the randomness of the FSO link quality induced by the instability of the UAV. Our simulation results reveal that buffering improves the performance of the system at the expense of an increased delay. We also show that the optimal position of the UAV is in general  different for BA and non-BA  UAV-based relay systems. 
	\item \textbf{Impact of weather conditions and MU density:} We show that the system performance and the optimal position of the UAV strongly depend on the atmospheric conditions of the backhaul channel and the density of the  MUs in the access channel. Our simulation results reveal  that  by optimal positioning of the UAV the impact of these system and channel parameters can be significantly reduced.  
\end{itemize}

The remainder of this paper is organized as follows. In Section \ref{sec2}, the considered system and channel models for UAV-based communication are presented. In Section \ref{Sec:analys}, the end-to-end  performance of  BA and non-BA UAV-based relaying systems is analyzed in terms of the ergodic sum rate, respectively.  Simulation results are provided in Section \ref{Sec_sim},  and conclusions are drawn in Section \ref{Sec_concl}. 

\textit{Notations:}
   In this paper, $(\cdot)^T$  and $(\cdot)^H$ denote the  transpose and  Hermitian transpose of a matrix, respectively.  $\mathbb{E}\{\cdot \}$, $*$, and $\lVert \cdot \rVert$ denote the expectation operator, the convolution operator, and the ${\ell}_2$-norm of a vector, respectively. $\mathcal{R}^+$ denotes the set of positive real numbers. $\mathbf{I}_{n}$ represents the  $n\times n$ identity matrix.  $\mathbf{x} \sim \mathcal{N} (\boldsymbol{\mu},\mathbf{\Gamma})$ and  $\mathbf{y} \sim \mathcal{CN} (\boldsymbol{\mu},\mathbf{\Gamma})$   indicate  that  $\mathbf{x}$ and $\mathbf{y}$ are respectively real and complex Gaussian random vectors with mean vector $\boldsymbol{\mu}$ and covariance matrix $\mathbf{\Gamma}$.   $x\sim \mathcal{U}(a,b)$ means that random variable (RV) $x$ is  uniformly distributed   in interval $[a, b]$. $y\sim\mathcal{H}(q,\omega)$  represents a Hoyt distributed RV $y$  with shape parameter  $q$ and spread factor $\omega$. $z\sim\text{lognormal}(\mu,\sigma^2)$ is used to indicate that  $z$ is a lognormal distributed RV where  $\mu$ and $\sigma^2$ are the  mean and variance in  $\text{dB}$.   Finally,
   $w\sim\mathrm{Nakagami}(m,\Omega)$ indicates that $w$ is a Nakagami distributed RV with shape parameter  $m$ and spread factor $\Omega$. 
\section{System and Channel Models}\label{sec2}
In this section, first we present the system model for the considered UAV-based relay network facilitating uplink communication between multiple MUs and a GS. Subsequently, we introduce the channel models for the MU-to-UAV RF access links and the UAV-to-GS FSO backhaul link.        
\subsection{System Model}
\begin{figure}
	\centering
	\includegraphics[width=0.7\textwidth]{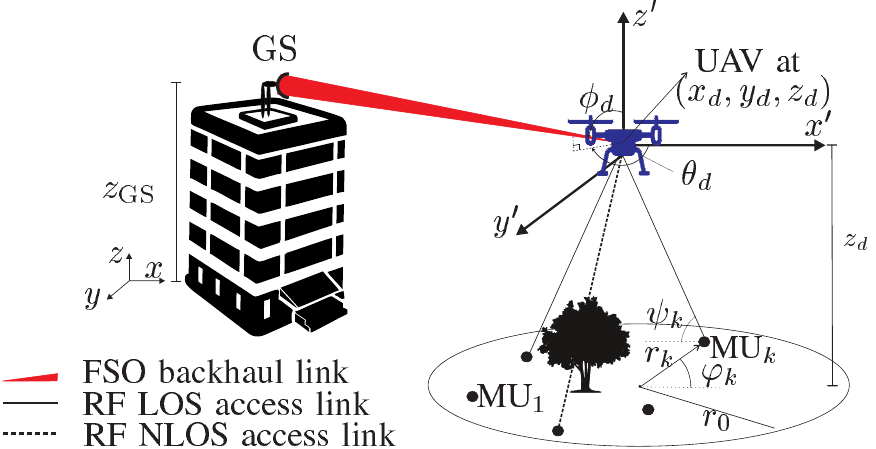}
	\caption{UAV-based communication system with FSO backhaul connection to the GS.}
	\label{Fig:SysMod}\vspace{-0.3cm}
\end{figure}
We consider an uplink UAV-based communication system, cf.~Fig.~\ref{Fig:SysMod},  with $K$ single-antenna MUs transmitting data over  RF links to a hovering UAV equipped with $N$ RF antennas and  a single aperture which relays  the received data over an FSO backhaul link to a GS equipped with a single PD. The position of the GS,  MUs, and  UAV are as follows. The GS is installed at height  $z_\mathrm{GS}$ of a  building  located in the origin of the Cartesian coordinate system $(x,y,z)$, i.e., the GS has coordinates $(0,0,z_\mathrm{GS})$. Moreover, the $K$ MUs are randomly distributed in a cell of radius $r_0$  centered at $(x_0,y_0,0)$. The random position of $\text{MU}_k, k\in \{1, 2, \dots, K\}$, is modeled by  a homogeneous Poisson point process, $\Phi$, with density $\lambda$, and is characterized by polar coordinates $ (r_k,\varphi_k)\in \Phi$, where $0\leq r_k \leq r_0$ is the radial distance from the cell center and $\varphi_k\in [0,2\pi]$ is the polar angle. Furthermore, the UAV is located at position $\mathbf{p}_d=(x_d,y_d,z_d)$ and its beam direction is determined by orientation variables $\mathbf{o}_d=(\theta_d,\phi_d)$. Consider  $(x',y',z')$ as a translation of the primary coordinate system, $(x,y,z)$, by vector $(x_d, y_d, z_d)$, cf.~Fig.~\ref{Fig:SysMod}.  Then,  $\phi_d$ is defined as the angle between the laser beam and the $z'$-axis and $\theta_d$ is the angle between the $x'$-axis and the projection of the beam onto the $x'-y'$ plane.    

Now, the uplink transmission can be divided into  two hops, the MUs-to-UAV hop and the UAV-to-GS hop.   The {MU}s are connected via RF access links to the UAV and by assuming frequency division multiple access (FDMA), each MU's signal is assigned to an orthogonal subchannel. 
Thus, the  signal received at the UAV from $\text{MU}_k$  is given by 
\begin{IEEEeqnarray}{rll}
\label{EQ:RFsystem}
\mathbf{y}_{k}=\mathbf{h}_{k} x_k+\mathbf{n}_{k},
\end{IEEEeqnarray}
where $\mathbf{h}_{k}=[h_{k,1}, \dots, h_{k,n}, \dots, h_{k,N}]^T$, $h_{k,n}$ is the flat fading channel coefficient from  $\text{MU}_k$ to  the $n$-th antenna of the UAV, and  $x_k$ is the transmit symbol of $\text{MU}_{k}$ with power $P=\mathbb{E}\{|x_k|^2 \}$. Moreover, $\mathbf{n}_{k} \sim \mathcal{CN} (\mathbf{0},\zeta^2 \mathbf{I}_{N})$ is  complex circularly symmetric additive white Gaussian noise.  Assuming  perfect channel state information (CSI) at the UAV, the elements of the received signal vector, $\mathbf{y}_{k}$, are  combined by  maximum ratio combining (MRC), and the resulting signal is decoded and forwarded to the GS via the FSO link. Furthermore, the UAV can transmit and receive simultaneously due to the mutual  orthogonality of the  RF and  FSO channels. We consider both non-BA and BA relaying at the UAV. In the former case, the UAV immediately forwards the data received from the MUs to the GS, whereas, in the latter case, the UAV can select to either receive and transmit the packets in the same time slot or to store them  in its buffer and  forward them in a later time slot when the FSO channel conditions are more favorable \cite{Nicola_FDBA}. Therefore, unlike non-BA relaying, BA relaying relaxes the constraint to transmit and receive according to a predetermined schedule and the UAV can select the best strategy (transmit, receive, or transmit and receive simultaneously) based on the  conditions of the  MUs-to-UAV and UAV-to-GS links. Hence, BA relaying can enhance the end-to-end achievable rate at the expense of an increased delay \cite{vahid_FSOrelay}, \cite{najafi_BA}. 

For the UAV-to-GS hop, the UAV maintains an FSO connection to the GS via a single laser aperture.  Next, assuming perfect CSI at  the GS and  an intensity modulation and direct detection (IM/DD) FSO system, the signal intensity received at the {GS}'s PD can be modeled as
\begin{IEEEeqnarray}{rll}
\label{EQ:FSOsystem}
\bar{y}=g \bar{x}+\bar{n},
\end{IEEEeqnarray}
where   $\bar{x}\in\mathcal{R}^+$ is the intensity modulated optical signal, $g$ is the scalar {FSO} channel coefficient, and $\bar{n}\sim\mathcal{N}(0,\rho^2)$ is the background Gaussian shot noise obtained after removing the ambient noise \cite{Lapid}.

\subsection{UAV-Based RF Channel Model}
The main difference between  UAV-based {RF} channels and the conventional channel models for satellite  and terrestrial mobile communications lies in the characteristics  of the LOS component. In particular, in satellite channels, the LOS path is almost always present, leading to Rician fading, whereas in urban mobile communication,  due to the large number of obstacles (e.g.~high rise buildings) in the environment, having an LOS path is  less probable and Rayleigh fading is expected \cite{Book_Alouini}. On the other hand, UAV-based communication is performed at altitudes that are between these two cases. Thus, the presence and absence of LOS links depends on the position of the UAV. This behavior was modeled in \cite{Hourani} and \cite{Hourani_opt_alt} via the probability of  attaining an LOS link between $\text{MU}_{k}$ and the UAV which is  given as follows 
\begin{IEEEeqnarray}{lll}\label{LOS_prob}
\mathrm{P}_{\mathrm{LOS}, k}=\frac{1}{1+C\exp(-B\psi_k)}\ ,
\end{IEEEeqnarray}
where $\psi_k=\frac{180}{\pi} \tan^{-1}(\frac{z_d}{r_{DM}})$ is the elevation angle between the UAV and  $\text{MU}_k$, $r_{DM}=\big((x_d-x_0-r_k\cos \phi_k)^2+(y_d-y_0-r_k\sin \phi_k)^2\big)^{1/2}$ is the radial distance between the UAV and  $\text{MU}_{k}$, and $C$ and $B$ are constants whose values depend on the environment (e.g., rural, urban, and high-rise areas).
Eq.~(\ref{LOS_prob})  suggests  that, at higher UAV operating altitudes, the {LOS} link is more likely to be present, which in turn comes at the expense of a higher path loss. Accordingly, the probability of having a non-LOS (NLOS) connection from the UAV to $\text{MU}_{k}$ is given by  $\mathrm{P}_{\mathrm{NLOS},k}=1-\mathrm{P}_{\mathrm{LOS},k}$.
Now,  based on the LOS probability, the RF channel coefficient, $h_{k,n}$, may be either  LOS or NLOS  and is given by \cite{Hourani, Rice_UAV}
\begin{IEEEeqnarray}{rll}\label{EQ:RFchannel}
h_{k,n}=
\begin{cases}
h^{p}_k h^{r}_{k,n} e^{j\Psi_{k,n}}, &\quad \mathrm{NLOS},\\
h^{p}_k h^{sr}_{k,n}, &\quad \mathrm{LOS},
\end{cases}
\end{IEEEeqnarray}
where $h^{p}_k$ is the free-space path loss, and $h^{r}_{k,n} e^{j\Psi_{k,n}}$  and $h^{sr}_{k,n}$ are the Rayleigh and  shadowed Rician fading coefficients, respectively. In particular, the path loss is given by $h^{p}_k=\frac{1}{c\sqrt{{r}_{DM}^2+z_d^2}}$, where  $c={f\over23.85}$ and  $f$ is the center operating frequency in $\text{MHz}$.
The NLOS scenario is characterized by the Rayleigh fading coefficient, $h^{r}_{k,n} e^{j\Psi_{k,n}}$, with power $\mathbb{E}\{\left(h^{r}_{k,n}\right)^2\}=\eta^2$ and  uniformly distributed phase, $\Psi\sim\mathcal{U}(0,2\pi)$. The probability density function (pdf)  of $h^{r}_{k,n}$, denoted by $f_{h^{r}_{k,n}}(x)$, is given by
\begin{IEEEeqnarray}{rll}\label{EQ:Rayleigh_dist}
f_{h^{r}_{k,n}}(x)={{x}\over{\eta^2}} e^{-{x^2}\over{2\eta^2}}.
\end{IEEEeqnarray}
Moreover, the distribution of $\varsigma=\sum\limits_{n=1}^N |h_{k,n}^{r}|^2$, which characterizes the combined signal after MRC at the UAV,  is given by a chi-distribution with $2N$ degrees of freedom, i.e., $\varsigma\sim\chi^2(2N)$.
In the LOS case, the channel is affected by both shadowing and small scale fading which is modeled via the Loo model \cite{New_sat}, \cite{Alouni_dronechannel}, \cite{UAV_lowelevation}. In this model, the  shadowed Rician fading coefficient is modeled as $h^{sr}_{k,n}\overset{\Delta}{=}h^{s}_k+h^{r}_{k,n} e^{j\Psi_{k,n}}$, where $h^{s}_k\sim\text{lognormal}(\mu,\sigma^2)$ is the lognormal shadowed LOS component which is added to the Rayleigh scattering component. Here, the log-normal shadowed component is identical across all RF antennas of the UAV, but the small scale Rayleigh fading  is independent across antennas.  Unfortunately, the combination of log-normal shadowing and Ricean fading  does not lend itself to a closed-form expression for the resulting distribution. Thus, as a widely accepted approximation \cite{New_sat}, \cite{Samimi}, the log-normal distribution is fitted to the Nakagami distribution  and   the shape and spread parameters of the Nakagami pdf are obtained via moment matching to the log-normal pdf as follows
\begin{IEEEeqnarray}{rll}\label{EQ:Nakag_Param}
q=\frac{1}{\exp(\frac{4\sigma^2}{\varepsilon^2})-1}, \quad
\omega=\exp\left({2\over\varepsilon}\left(\mu+\frac{\sigma^2}{\varepsilon}\right)\right),
\end{IEEEeqnarray}
where $\varepsilon=\frac{20}{\ln(10)}$ and thus, $h^{s}_k\sim\text{Nakagami}(q,\omega)$. Based on this approximation, the  shadowed-Rician pdf  is modeled in \cite{New_sat}. In the following Lemma, we characterize the distribution of $\tau=\sum\limits_{n=1}^N |h_{k,n}^{sr}|^2$.

\begin{lem}\label{Lemma3}
	For the proposed Nakagami approximation of $h_k^s$, the distribution of $\tau=\sum\limits_{n=1}^N |h_{k,n}^{sr}|^2$  is given by
	\begin{IEEEeqnarray}{rll}\label{dist_tau}
		&f_{{\tau}}\left(x\right)= \frac{\left(\frac{N\omega}{2q\eta^2}+1\right)^{-q}x^{N-1} }{2^{N}\eta^{2N}\left(N-1\right)!e^{\frac{x}{2\eta^2}}} \,_1\mathcal{F}_1\left(q,N;\frac{x}{2\eta^2+\frac{4\eta^4 q}{N\omega}}\right),\qquad
	\end{IEEEeqnarray}
where $_1 F_1(\cdot, \cdot; \cdot)$ is the confluent hypergeometric function.
\end{lem}
\begin{IEEEproof}
	The proof is given in Appendix \ref{App5}.
\end{IEEEproof}

Thus, the pdf of  $\lVert{\mathbf{{h}}}_k\rVert^2=\sum\limits_{n=1}^N |h_{k,n}|^2$   is given by
\begin{IEEEeqnarray}{rll}\label{EQ:Totchannel_dist}
f_{\lVert{\mathbf{{h}}}_k\rVert^2}(\hspace*{-.5mm}h)=\frac{1}{|h^{p}_k|^2}\left( f_{\tau}\left(\frac{h}{|h^{p}_k|^2}\right) \mathrm{P}_{\mathrm{NLOS},k}+f_{\varsigma}\hspace*{-1mm}\left(\hspace*{-1mm}\frac{h}{|h^{p}_k|^2}\hspace*{-1mm}\right)\hspace*{-1mm}\mathrm{P}_{\mathrm{LOS},k}\hspace*{-1mm}\right)\hspace*{-1mm}.\nonumber\\
\end{IEEEeqnarray}

\subsection{UAV-Based FSO Channel Model}
The UAV-based FSO channel differs from conventional FSO channels with  fixed transceivers mounted on top of buildings in the following two aspects. First, in contrast to fixed transceivers, the FSO beam of the UAV is not necessarily orthogonal to the PD plane.  Second, the instability of the UAV, i.e., the random vibrations of the UAV,  introduces a random power loss.
Taking these effects  into account, the point-to-point UAV-based {FSO} channel, $g$, can be modeled as follows \cite{ICC, Stat_Marzieh} 
\begin{IEEEeqnarray}{lll}\label{FSO_Channel}
g=R_s g_p g_a g_g,
\end{IEEEeqnarray}
where $R_s$, $g_p$, $g_a$, and $g_g$ represent the responsivity of the PD, the atmospheric loss, the atmospheric turbulence, and the geometric and misalignment loss (GML), respectively. 
\subsubsection{Atmospheric Loss}
The atmospheric loss, $g_p$, is due to scattering and absorption of the laser beam by atmospheric particles and is given by \cite{ICC}
\begin{IEEEeqnarray}{lll}\label{g_p}
g_p=10^{-\frac{\kappa L}{10}},
\end{IEEEeqnarray}
where $\kappa$ is the attenuation factor, whose value depends on the weather conditions (e.g.\  clear, foggy), and  $L=\sqrt{x_d^2+y_d^2+(z_d-z_\mathrm{GS})^2}$ is the distance between the UAV and the GS.
\subsubsection{Atmospheric Turbulence}\label{subsec_GGassum}
The atmospheric turbulence, $g_a$,  is caused by variations of the refractive index in different layers of the atmosphere due to fluctuations in pressure and temperature. As a universal model that considers both small and large  scale irradiance fluctuations, the Gamma-Gamma (GG) distribution, $g_a\sim \mathcal{GG}(\alpha,\beta)$, is considered. Here, $\alpha$ and $\beta$ are the small and large scale turbulence parameters, respectively, and are given by \cite{GG_Murat}
\begin{IEEEeqnarray}{lll}\label{alphabeta}
&\alpha =\left[\exp\left ({{0.49\sigma_{R}^{2}}\over {\left (1+0.18\iota^2+0.56\sigma_{R}^{\frac{12}{5}}\right) ^{\frac{7}{6}}}}\right)\hspace{-2mm}-1\right] ^{-1}\hspace{-5mm},\nonumber \\ &\beta =\left[\exp \left ({{0.51\sigma _{R}^{2}\left(1+0.69\sigma_R^{\frac{12}{5}}\right)^{-\frac{5}{6}}}\over {\left (1+0.9\iota^2+0.62\iota^2\sigma _{R}^{\frac{12}{5}}\right) ^{\frac{5}{6}}}}\right)\hspace{-2mm}-1\right]^{-1}\hspace{-5mm},\quad\quad
\end{IEEEeqnarray}
where $\iota=\left(k(2a)^2/4L\right)^{1/2}$, $\sigma_R^2= 1.23C_{n}^{2}k^{{7}/{6}}L^{{11}/{6}}$ is the Ryotov variance, $C_{n}^{2}$ represents  the index of refraction structure parameter,  $k$ denotes the wave number, and  $a$ is the radius of the PD. The Hufnagle-Valley (H-V) model suggests that $C_n^2$ decreases with increasing height of the UAV (up to 3 km) and is given by $C_n^2\approx C_{n}^{2}(0)\exp\left({-z_d\over 100}\right)$,
 where $C_{n}^{2}(0)$ is the ground level refraction structure parameter  \cite{FSO_Drone_Alouni}, \cite{Andrews_book}. Thus, the Ryotov variance and accordingly the scintillation variance, $\mathrm{var}\{g_a\}=\frac{1}{\alpha}+\frac{1}{\beta}+\frac{1}{\alpha\beta}$, depend on both the  distance between the laser aperture and the {PD},  $L$, and  the UAV operating height, $z_d$. Now, for conventional applications where UAVs are used to build temporary networks,  short distances between the UAV and the GS, e.g., $L\leq 600\ \text{m}$, are  expected. In this operating range and  typical heights of  $z_d>30\ \text{m}$,  the impact of scintillation becomes very weak even in clear weather condition, i.e., $\mathrm{var}\{g_a\}\leq 0.15$, and hence for our analysis in Section \ref{Sec:analys}, we ignore the effect of turbulence, i.e., $g_a=\mathcal{E}\{g_a\}=1$ is assumed. Then, in Section \ref{Sec_sim}, we use simulations to investigate the impact of this simplifying assumption on the  end-to-end achievable rate of the system.

\subsubsection{GML}
The GML, $g_g$,  comprises  the geometric loss due to the beam spread along the propagation path and the misalignment loss due to the random fluctuations of the position and orientation of the UAV. These random  fluctuations are caused by  different phenomena, including  random air fluctuations  around the UAV,  internal vibrations of the UAV, and  tracking errors, see  \cite{Stat_Marzieh} for a detailed discussion. Also, for the UAV-based FSO channel,  the  positioning of the UAV may lead to non-orthogonality between the laser beam and the {PD} plane, which in turn introduces an additional geometric loss. Taking the aforementioned effects into account, the GML for UAV-based  FSO channels can be modeled as follows \cite{ICC}  
\begin{IEEEeqnarray}{rll}\label{g_g}
g_g=A_0 \exp\left(- \frac{2u^2}{k_{g}w^2}\right),
\end{IEEEeqnarray}
where $u$ is the misalignment factor, $w$ is the beam width at distance $L$, $A_0=\text{erf}(\nu_\mathrm{min})\text{erf}(\nu_\mathrm{max})$, $k_g= \frac{k_\mathrm{min}+k_\mathrm{max}}{2}$, $k_{\mathrm{min}}=\frac{\sqrt{\pi}\mathrm{erf}(\nu_\mathrm{min})}{2\nu_\mathrm{min}\exp(-\nu^2_\mathrm{min})}$, $\nu_\mathrm{min}=\sqrt{\frac{\pi}{2}}\frac{a}{w}$, $k_{\mathrm{max}}=\frac{\sqrt{\pi}\mathrm{erf}(\nu_\mathrm{max})}{2\sin^2(\phi_d)\cos^2(\theta_d)\nu_\mathrm{max}\exp(-\nu^2_\mathrm{max})}$, and $\nu_\mathrm{max}=|\sin(\phi_d)\cos(\theta_d)|\nu_\mathrm{min}$. Here, $\text{erf}(x)=\frac{1}{\sqrt{\pi}}\int_{-x}^{x}\exp(-t^2) \mathrm{d}t$ is the error function.

Assuming  perfect tracking, i.e., $\mathbb{E}\{u\}=0$, the random variations of the position and orientation of the UAV can be characterized  by a Hoyt distributed misalignment factor, i.e., $u\sim \mathcal{H}(m,\Omega)$, and accordingly, the pdf of the GML is given by \cite{ICC}
\begin{IEEEeqnarray}{rll}\label{g_g_dist}
f_{g_g}(g_g)&=\frac{\varrho}{A_0}\left(\frac{g_g}{A_0} \right)^{\frac{(1+m^2)\varrho}{2m}-1}\nonumber\\
&\times{I}_0\left( -\frac{(1-m^2)\varrho}{2m} \ln \left(\frac{g_g}{A_0}\right) \right), \hspace{0.5mm} 0\leq g_g\leq A_0,\quad
\end{IEEEeqnarray}
where ${I}_0(\cdot)$ is the zero-order modified Bessel function of the first kind, $\varrho=\frac{(1+m^2)k_{g} w^2}{4m\Omega}$, $m=\sqrt{\frac{\min\{\lambda_1,\lambda_2\}}{\max\{\lambda_1,\lambda_2\}}}$, $\Omega=\lambda_1+\lambda_2\ $, and  $\lambda_1$ and  $\lambda_2$ are the eigenvalues of matrix
\begin{IEEEeqnarray}{rll}\label{matrix} \mathbf{\Sigma}=\begin{bmatrix} \sigma_{y_d}^{2}+c_{1}^{2}\sigma_{x_d}^{2}+c_{2}^{2}\sigma_{\theta_d}^{2}&c_{1}c_{5}\sigma_{x_d}^{2}+c_{2}c_{4}\sigma_{\theta_d}^{2}\\ c_{1}c_{5}\sigma_{x_d}^{2}+c_{2}c_{4}\sigma_{\theta_d}^{2}&\sigma_{z_d}^{2}+c_{5}^{2}\sigma_{x_d}^{2}+c_{4}^{2}\sigma_{\theta_d}^{2}+c_{3}^{2}\sigma_{\phi_d}^{2}
 \end{bmatrix}.\nonumber\\\quad
\end{IEEEeqnarray}
Here, $\sigma^2_{i} , i\in\{x_d,y_d,z_d,\theta_d,\phi_d\}$, denotes the variance of the random fluctuations of the UAV along the position and orientation variables, $c_1=-\tan(\theta_d)$, $c_2=-\frac{x_d}{\cos^2(\theta_d)}$, $c_3=\frac{x_d}{\sin^2(\phi_d)\cos(\theta_d)}$, $c_4=-\frac{x_d \cot(\phi_d)\tan(\theta_d)}{\cos(\theta_d)}$, and $c_5=-\frac{\cot(\phi_d)}{\cos(\theta_d)}$.

 Next, given (\ref{FSO_Channel}) and (\ref{g_g_dist}), the pdf of the FSO channel, disregarding the atmospheric turbulence, can be modeled as
\begin{IEEEeqnarray}{rll}\label{FSO_ditr}
	f_{g}(g)=\frac{1}{R_s  g_p}f_{g_g}\left(\frac{g}{R_s  g_p}\right), \quad 0\leq g\leq R_s  g_pA_0.\end{IEEEeqnarray}

\begin{remk}\label{remk1}
To shed some light on the impact of the various system parameters on the distribution of the misalignment factor $u$, we consider  special cases.	Let us assume $\sigma_{x_d}=\sigma_{y_d}=\sigma_{z_d}=\sigma_{p}$ and $\sigma_{\theta_d}=\sigma_{\phi_d}=\sigma_{o}$.  If the UAV flies in the $x$\nobreakdash--$z$ plane, i.e., $y_d=\theta_d=0$, then we obtain  $\lambda_1=\sigma_p^2+x_d^2 \sigma_o^2$, $\lambda_2=\sigma_p^2(1+\cot^2\phi_d)+\frac{x_d^2}{\sin^2\phi_d}\sigma_o^2$, $m={\lambda_1\over \lambda_2}$, and $\Omega=\lambda_1+\lambda_2$. Under this assumption,	we  consider the following special cases for $u\sim \mathcal{H}(m,\Omega)$:
	\begin{itemize}[leftmargin=3mm]
		\item   UAV flies along the $z$-axis and $x_d=x_0$: By increasing $|z_d-z_\mathrm{GS}|$,  $m$ reduces and  $\Omega$  increases.
		\item   UAV flies along the $x$-axis and $z_d=z_\mathrm{GS}$: Now,  $\phi_d=\pi/2$ and $m=1$  and  decreasing $x_d$ reduces $\Omega$.
		\item 	UAV hovers in front of the GS: $y_d=0$, $z_d=z_\mathrm{GS}$, $\theta_d=0$, and $\phi_d=\frac{\pi}{2}$, then we obtain  $m=1$, and $\Omega=2(\sigma_p^2+x_d^2 \sigma_o^2)$. In this case,  the FSO beam is orthogonal to the PD plane and  the misalignment factor $u$ is  Rayleigh distributed.
	\end{itemize}
\end{remk}
In summary, the models for  both  the {RF} channel and the {FSO} channel of  UAV-based relay networks differ substantially from the corresponding models for conventional relay networks without UAVs. In particular, the positioning  of the UAV affects the path loss and  LOS characteristics of the RF access links and the atmospheric loss of the FSO backhaul  channel. Furthermore, the instability of the UAV impacts the GML of the FSO backhaul channel.  In the following, we study the impact of these effects on the end-to-end  achievable rate of the system.

\section{End-to-End Ergodic Sum Rate Analysis}\label{Sec:analys}
In this section, the ergodic sum rate, $\mathbb{E}\{C_{\mathrm{sum}}\}$, is adopted as a metric for analyzing the end-to-end system performance.
Assuming DF relaying at the UAV, the end-to-end achievable sum rate is restricted to the rate of the weaker of the two involved links as a consequence of the max-flow min-cut theorem \cite{ElGamal}. In particular, for  non-{BA} relaying, the achievable sum rate depends on  the instantaneous fading states of both  hops \cite{vahid_TWC_partII}. Thus, the non-BA ergodic sum rate, denoted by $\bar{C}^\mathrm{NB}_{\mathrm{sum}}$, is given by 
\begin{IEEEeqnarray}{rll}\label{non-BA_rate}
\bar{C}^\mathrm{NB}_{\mathrm{sum}}=\mathbb{E}\{C_{\mathrm{sum}}\}=\mathbb{E}\{ \min\left( C^{\mathrm{RF}},C^{\mathrm{FSO}}\right) \},
\end{IEEEeqnarray}
where $C^{\mathrm{RF}}$ and $C^{\mathrm{FSO}}$ are the instantaneous achievable rates of the {RF} and {FSO} links, respectively.
In the BA scenario, the relay is equipped with buffers to store the  data received from the MUs and to transmit it when the FSO channel is in a favorable state \cite{vahid_TWC_partI}. Here, for unlimited  buffer sizes, the BA ergodic sum rate is given by
\begin{IEEEeqnarray}{rll}\label{BA_rate}
\bar{C}_{\mathrm{sum}}^{\mathrm{BA}}=\min\left(\mathbb{E}\{  C^{\mathrm{RF}}\},\mathbb{E}\{C^{\mathrm{FSO}} \}\right).
\end{IEEEeqnarray}
\begin{remk} 
	Although, we assume an unlimited buffer size in (\ref{BA_rate}), in practice, the buffer size is finite and hence, (\ref{BA_rate}) constitutes a performance upper bound for practical BA relaying systems. However, in \cite{vahid_TWC_partI} and \cite{vahid_TWC_partII}, it has been shown that the performance of BA relays with sufficiently large  buffer sizes   closely approaches the upper bound for unlimited buffer size. 
\end{remk}
\begin{remk}
 Exploiting Jensen's inequality for concave $\min$ function, $\mathbb{E}\{f(x)\}\leq f(\mathbb{E}\{x\})$, we can relate the ergodic sum rates for  non-{BA} and {BA} relaying  as follows  
\begin{IEEEeqnarray}{rll}\label{BA and NBA}
\bar{C}^\mathrm{NB}_{\mathrm{sum}}\leq \bar{C}_{\mathrm{sum}}^{\mathrm{BA}}.
\end{IEEEeqnarray}
Hence, the  ergodic sum rate achieved with BA relaying is an upper bound for the ergodic sum rate of the non-{BA} case which suggests that buffering data is advantageous  for achieving a high ergodic sum rate for the  end-to-end system. Nevertheless, this gain comes at the expense of a higher end-to-end delay. Therefore, BA relaying is suitable for delay-tolerant applications. 
\end{remk}
Next, we analyze the ergodic rate  for  non-BA (\ref{non-BA_rate}) and  BA (\ref{BA_rate})  relay UAVs for the UAV-based mixed RF-FSO channel model presented in Section \ref{sec2} .
\subsection{Ergodic Rate Analysis for BA Relay UAV}
The achievable rate for a BA relay UAV depends only on the individual ergodic rates of the RF and FSO channels, i.e., $\mathbb{E}\{  C^{\mathrm{RF}}\}$ and $\mathbb{E}\{C^{\mathrm{FSO}} \}$ in (\ref{BA_rate}). Therefore, in the following, we analyze these ergodic rates separately.

\subsubsection{Achievable Ergodic Sum Rate of the RF Channel}
Given that the MUs employ orthogonal subchannels, the instantaneous rate of the RF channel, $C^{\mathrm{RF}}$, can be written as a summation of all MUs' rates and is given by
\begin{IEEEeqnarray}{rll}\label{inst_RF_rate}
C^{\mathrm{RF}}=W_{\mathrm{sub}}^{\mathrm{RF}} \sum\limits_{(r_k,\phi_k)\in \Phi}R_k^{\mathrm{RF}},
\end{IEEEeqnarray}
where ${W}_{\mathrm{sub}}^{\mathrm{RF}}$ is the subchannel bandwidth and $R_{k}^{\mathrm{RF}}=\log_2(1+\frac{P}{\zeta^2}\lVert \mathbf{h}_k\rVert^2)$ is the achievable rate of  $\text{MU}_k$. 
Now, the RF ergodic rate is determined by averaging over the  random  fluctuations of the shadowed Rician and  Rayleigh fading in the RF channel and the random MU positions. Taking these effects into account, the ergodic sum rate of the RF channel, denoted by $\bar{C}^\mathrm{RF}$,  is given in the following theorem. 

\begin{theo}\label{Theorem1}
The ergodic sum rate of the RF channel is given as follows
\iftoggle{OneColumn}{\begin{IEEEeqnarray}{rll}\label{theo1}
		\bar{C}^\mathrm{RF}&=\mathbb{E}_{h,\Phi}\{C^{\mathrm{RF}}\}= \frac{\pi^3\lambda r_0^2 }{4H M} W_{\mathrm{sub}}^{\mathrm{RF}}\sum_{i=1}^{H} \sum_{j=1}^{M} \sqrt{(1-x_i^2)(1-y_j^2)}(x_i+1)  \big[ \mathbb{E}_{h_{\mathrm{r}}}\{R^{\mathrm{RF}}\}+\nonumber\\
		&+\left( \mathbb{E}_{h_{\mathrm{sr}}}\{R^{\mathrm{RF}}\}- \mathbb{E}_{h_{\mathrm{r}}}\{R^{\mathrm{RF}}\}\right)  P_{\mathrm{LOS}}\big]+E_H+E_M,
\end{IEEEeqnarray}}{\begin{IEEEeqnarray}{rll}\label{theo1}
\bar{C}^\mathrm{RF}&=\frac{\pi^3\lambda r_0^2 W_{\mathrm{sub}}^{\mathrm{RF}}}{4H M}\sum_{i=1}^{H}  \sum_{j=1}^{M} \sqrt{(1-x_i^2)(1-y_j^2)}(x_i+1)  \nonumber\\
&\times\big[ \mathbb{E}_{h_{\mathrm{r}}}\{R^{\mathrm{RF}}\}+\left( \mathbb{E}_{h_{\mathrm{sr}}}\{R^{\mathrm{RF}}\}- \mathbb{E}_{h_{\mathrm{r}}}\{R^{\mathrm{RF}}\}\right) P_{\mathrm{LOS}}\big]\nonumber\\
&+E_H+E_M,\qquad
\end{IEEEeqnarray}}
 where $R^{\mathrm{RF}}=\log_2(1+\gamma\lVert{\mathbf{\tilde{h}}}_k\rVert^2) $, ${\mathbf{\tilde{h}}}_k=\frac{1}{h_k^p}{\mathbf{h}}_k$, $\gamma=\frac{{P}}{ {\zeta^2}c^2(z_d^2+\tilde{r}_{\mathrm{DM}}^2)} $, $P_{\mathrm{LOS}}=P_{\mathrm{LOS},k}(\tilde{\psi}_k)$ , $\tilde{\psi}_k=\frac{180}{\pi}\tan^{-1}(\frac{z_d}{\tilde{r}_{\mathrm{DM}}})$,  $\tilde{r}_{\mathrm{DM}}^2=(x_d-x_{0}+\frac{r_0}{2}(x_i+1)\cos(\pi y_j))^2+(y_d-y_{0}+\frac{r_0}{2}(x_i+1)\sin(\pi y_j))^2$,  $x_{i}=\cos\left(\frac{2i-1}{2H}\pi\right)$, $y_{j}=\cos\left(\frac{2j-1}{2M}\pi\right)$. Additionally, the ergodic rates for Rayleigh  and shadowed-Rician fading are respectively given as follows 
\iftoggle{OneColumn}{\begin{IEEEeqnarray}{rll}
		\mathbb{E}_{h_{\mathrm{r}}}\{R^{\mathrm{RF}}\}&=\sum_{\ell=0}^{N-1}\frac{1}{(2\eta^2\gamma)^{\ell}}\  e^{\frac{1}{2\eta^2\gamma}}\  \Gamma\left(-\ell,\frac{1}{2\eta^2\gamma}\right),\label{Rate_r,sr}\\
		\mathbb{E}_{h_{\mathrm{sr}}} \{R_k^{\mathrm{RF}}\}&=\frac{\left(\frac{N\omega}{2q\eta^2}+1\right)^{-q}e^{-1/2\eta^2\gamma}}{2^{N}\eta^{2N}}\sum\limits_{n=0}^\infty\frac{(q)_n}{n!\left(2\eta^2+\frac{4\eta^4q}{N\omega}\right)^n\gamma^{n+N}} \nonumber\\
		&\times\sum\limits_{\ell=1}^{n+N} \Gamma\left(\ell-n-N,\frac{1}{2\eta^2\gamma}\right)\left(2\eta^2\gamma\right)^\ell,\quad\label{Rate_sr}
\end{IEEEeqnarray}}{
\begin{IEEEeqnarray}{rll}
\mathbb{E}_{h_{\mathrm{r}}}\{R^{\mathrm{RF}}\}&=\sum_{\ell=0}^{N-1}\frac{1}{(2\eta^2\gamma)^{\ell}}\  e^{\frac{1}{2\eta^2\gamma}}\  \Gamma\left(-\ell,\frac{1}{2\eta^2\gamma}\right),\label{Rate_r,sr}\\
\mathbb{E}_{h_{\mathrm{sr}}} \{R_k^{\mathrm{RF}}\}&=\frac{\left(\frac{N\omega}{2q\eta^2}+1\right)^{-q}e^{-1/2\eta^2\gamma}}{2^{N}\eta^{2N}}\nonumber\\
&\times\sum\limits_{n=0}^\infty\frac{(q)_n}{n!\left(2\eta^2+\frac{4\eta^4q}{N\omega}\right)^n\gamma^{n+N}} \nonumber\\
&\times\sum\limits_{\ell=1}^{n+N} \Gamma\left(\ell-n-N,\frac{1}{2\eta^2\gamma}\right)\left(2\eta^2\gamma\right)^\ell,\quad\label{Rate_sr}
\end{IEEEeqnarray}}
where  $\Gamma(\cdot,\cdot)$ and $(x)_n$ denote incomplete Gamma function and  the (rising) Pochhammer symbol, respectively \cite{integral}.
\end{theo}
\begin{IEEEproof}
The proof is given in Appendix \ref{App1}.
\end{IEEEproof}

For the error terms, we have $E_H, E_M\rightarrow0$ as $H$ and $M$ increase. In Section \ref{Sec_sim}, we will show that even for small values of $H$ and $M$ (e.g., $H=M=10$),  (\ref{theo1}) yields accurate results. Theorem \ref{Theorem1} explicitly reveals the dependence of the ergodic sum rate of the RF channel on the position of the UAV via parameters  $\gamma$ and $P_{\mathrm{LOS}}$. In particular, by moving the UAV upwards (larger $z_d$) or towards the GS (smaller $x_d$ or $y_d$),  $\gamma$  and accordingly both ergodic sum rate terms, $\mathbb{E}_{h_{\mathrm{sr}}}\{R^{\mathrm{RF}}\} $ and $\mathbb{E}_{h_{\mathrm{r}}}\{R^{\mathrm{RF}}\}$, decrease. Given the LOS path in $h_{sr}$, for the same multipath power $\eta^2$, the ergodic sum rate for shadowed Rician fading is always larger than that for Rayleigh  fading, and  the term $\mathbb{E}_{h_{\mathrm{sr}}}\{R^{\mathrm{RF}}\}-\mathbb{E}_{h_{\mathrm{r}}}\{R^{\mathrm{RF}}\}$  decreases by moving the UAV further from the cell center. On the other hand, the LOS probability, $P_{\mathrm{LOS}}$, increases for larger $z_d$ and decreases for smaller $x_d$ and $y_d$. Therefore, the ergodic sum rate of the RF channel always degrades if the UAV moves away from the cell center towards the GS (smaller $x_d$ or $y_d$), but for vertical movement of the UAV (larger $z_d$), there is  a trade-off between the LOS probability and the path loss. Thus, the positioning of the UAV  plays an important role for the achievable rate of the RF channel. 
\subsubsection{Achievable Ergodic  Rate of the FSO Channel}
In the second  hop,  for an average power constraint, $\bar{p}$,  the achievable rate of an {IM/DD} FSO system is given by  \cite{Lapid}
\begin{IEEEeqnarray}{rll}\label{FSO_cap_inst}
C^{\mathrm{FSO}}= \frac{1}{2}W^\mathrm{FSO}\log_2\left(1+\frac{e\bar{p}^2}{2\pi\rho^2} g^2\right),
\end{IEEEeqnarray}
where $W^\mathrm{FSO}$ denotes the {FSO} bandwidth. Unfortunately, the ergodic rate of the FSO system cannot be computed in closed form for the entire range of  signal-to-noise ratios (SNRs). In fact, the expected value of (\ref{FSO_cap_inst}) w.r.t. the squared Hoyt variable, $u^2$, i.e., $\bar{C}^\mathrm{FSO}=\frac{W^\mathrm{FSO}(1+m^2)}{4m\Omega} $ $\times\int\limits_0^\infty \log_2\left(\!1+ \bar{\gamma}^2  e^{\!-\frac{4x}{k_gw^2}}\right) e^{-\frac{(1+m^2)^2x}{4m^2\Omega}}I_0\left(\frac{(1-m^4)x}{4m^2\Omega}\right)\mathrm{d}x$, does not have a closed-form solution and can only be obtained numerically. Nevertheless, the following theorem presents the ergodic rate for low and high SNRs.
\begin{theo}\label{Theorem2}
The ergodic rate of the FSO system  for the low and high SNR regimes, i.e., $\bar{\gamma}< 1$ and $\bar{\gamma}\gg 1$, respectively, is given  by
\begin{IEEEeqnarray}{rll}\label{Theo2}
\bar{C}^\mathrm{FSO}_\mathrm{low} &=\frac{W^\mathrm{FSO}}{\ln(2)} \sum_{\ell=1}^{\infty}\frac{(-1)^{\ell+1}\left(\bar{\gamma}\right)^{2\ell}}{\ell\sqrt{\frac{(4\ell)^2}{\varrho^2}+4\left(1+\frac{8\ell\Omega}{k_gw^2}\right)}},\qquad\IEEEyesnumber\IEEEyessubnumber\label{theo2}\\
\bar{C}^\mathrm{FSO}_\mathrm{high}&=\frac{W^\mathrm{FSO}}{2}\left(\log_2(\bar{\gamma}^2)-\frac{4\Omega}{\ln(2) k_gw^2}\right),\qquad\IEEEyessubnumber\label{theo2_II}
\end{IEEEeqnarray}
where $\bar{\gamma}=R_sg_pA_0\left(\frac{e\bar{p}^2}{2\pi\rho^2}\right)^{1\over 2}$.
\end{theo}
\begin{IEEEproof}
Please refer to Appendix \ref{App2}.
\end{IEEEproof}

As will be shown in Section \ref{Sec_sim}, (\ref{theo2})  yields an accurate approximation even if the number of summation terms is limited to only $5$. Here,   variables $w, \Omega, \varrho,g_p$, and $A_0$ are dependent on the position and orientation parameters of the UAV, namely $\mathbf{p}_d$ and $\mathbf{o}_d$.
Theorem \ref{Theorem2}  reveals that the ergodic rate of the FSO channel depends on the positioning of the UAV via  $g_p$, $w$, $A_0$ and on the instability of the UAV via Hoyt parameters $m$ and $\Omega$.  
 \begin{remk}\label{Remark 4}
To gain some insights, let us  consider the case where the UAV changes only its altitude and flies along the $x$-axis. When  the UAV is  located at the same height as the GS, the beam is orthogonal to the PD plane. Recall from Remark \ref{remk1}  that in this case, $m=1$, $\Omega$  assumes its minimum value and since, the beam has the maximum possible  footprint on the PD, $A_0$ takes its maximum value. Then, if the UAV moves to higher or lower altitudes than the GS, $m$ and $A_0$ decrease, $\Omega$ increases, and accordingly, the GML increases. Moreover, due to the larger distance between the UAV and the GS, the additional  atmospheric loss increases which further deteriorates the FSO channel. Hence,   $\bar{\gamma}$ decreases which in turn degrades the ergodic rate of the FSO channel. Thus,  the UAV's position  and its (in)stability   crucially affect the ergodic rate of the FSO channel via $\bar{\gamma}$ and parameters $m$ and $\Omega$, respectively.
 \end{remk}

Finally, the {BA} ergodic sum rate in (\ref{BA_rate}) is given by 
\begin{IEEEeqnarray}{rll}\label{BA_result}
\bar{C}_{\mathrm{low}}^{\mathrm{BA}}&=
\min\left(\bar{C}^\mathrm{RF},\bar{C}^\mathrm{FSO}_\mathrm{low} \right),  \quad  \bar{\gamma}< 1,\qquad\IEEEyesnumber\IEEEyessubnumber\label{BA_result_low}\\
\bar{C}_{\mathrm{high}}^{\mathrm{BA}}&=
\min\left(\bar{C}^\mathrm{RF},\bar{C}^\mathrm{FSO}_\mathrm{high} \right), \quad  \bar{\gamma}\gg 1,\qquad\IEEEyessubnumber\label{BA_result_high}
\end{IEEEeqnarray}
where $\bar{C}^\mathrm{RF}$, $\bar{C}^\mathrm{FSO}_\mathrm{low}$, and $\bar{C}^\mathrm{FSO}_\mathrm{high}$  are given by (\ref{theo1}), (\ref{theo2}), and (\ref{theo2_II}), respectively. Eq.~(\ref{BA_result})  illustrates  the inherent trade-off between the ergodic rates of the {RF} and {FSO} channels  which are  dependent on the position and the instability of the hovering UAV. The position of the UAV affects the {LOS} probability,  path loss, and  geometric loss of the RF and FSO channels and the instability of the UAV influences  the misalignment loss of the FSO channel. 
Considering these effects, in Section \ref{Sec_sim}, the expression in (\ref{BA_result}) is  employed to optimize the positioning of the BA UAV for maximization of the end-to-end system performance.  

\subsection{ Ergodic Rate Analysis for  Non-BA Relay UAV}
In this section, we assume that the UAV  is not equipped with a buffer and the instantaneous rate in each channel hop determines the  system's end-to-end  rate. The non-BA ergodic rate in (\ref{non-BA_rate}) can be written as \cite{Random_proces}
\begin{IEEEeqnarray}{rll}\label{nonBA_in CDFform}
\bar{C}^\mathrm{NB}_{\mathrm{sum}}=\int_{t>0} \mathrm{Pr}(x>t)\ \mathrm{d}t=\int_{0}^\infty (1-F_x(t))\ \mathrm{d}t,
\end{IEEEeqnarray}
where $x=\min\left( C^{\mathrm{RF}},C^{\mathrm{FSO}}\right)$. Here, the cumulative distribution function (CDF), $F_x(x)$,  depends on the {CDF} of both the {RF} and {FSO} channels as follows
\begin{IEEEeqnarray}{rll}\label{CDF_min}
F_x(x)=1-(1-F_{C^{\mathrm{RF}}}(x))(1-F_{C^{\mathrm{FSO}}}(x)).
\end{IEEEeqnarray}
Here, $F_{C^{\mathrm{RF}}}(x)=F_{R_1^{\mathrm{RF}}}(\tilde{x})*\dots *F_{R_K^{\mathrm{RF}}}(\tilde{x})$, where  $F_{R_k^{\mathrm{RF}}}(\tilde{x})$ denotes the CDF of the sum rate of $\text{MU}_k$ and $\tilde{x}=\frac{x}{W_{\mathrm{sub}}^{\mathrm{RF}}}$. Particularly, $F_{R_k^{\mathrm{RF}}}(\tilde{x})$  is  the summation of the CDFs of the sum of a squared  shadowed Rician RV and a squared  Rayleigh RV. Thus,  $F_{C^{\mathrm{RF}}}(u)$  does not lend itself to a closed-form expression. To cope with this issue, the following lemma is proposed.
\begin{lem}\label{Lemma2}
Assuming the number of  MUs is sufficiently large, $\lim\limits_{K\rightarrow \infty} {C}^{\mathrm{RF}}=\bar{C}^{\mathrm{RF}}$,  then  the  ergodic rate for the non-BA relay UAV is given by
\begin{IEEEeqnarray}{rll}\label{lem2}
\lim\limits&_{K\rightarrow \infty}\bar{C}^\mathrm{NB}_{\mathrm{sum}}=\nonumber\\
&\mathbb{E}_g\{\!C^{\mathrm{FSO}}|C^{\mathrm{FSO}}\leq \bar{C}^\mathrm{RF}\!\}+\bar{C}^{\mathrm{RF}}(1-F_{C^{\mathrm{FSO}}}(\bar{C}^{\mathrm{RF}})).\qquad
\end{IEEEeqnarray}
\begin{IEEEproof}
Please refer to Appendix \ref{App3}.
\end{IEEEproof} 
\end{lem}

In Section \ref{Sec_sim}, we provide simulation results to confirm that  even for comparatively small numbers of  MUs (e.g., $\bar{K}=\mathbb{E}\{K\}\geq50$), (\ref{lem2}) is an accurate approximation.
Based on (\ref{lem2}), two terms are needed to analyze the non-BA ergodic rate, namely $F_{C^{\mathrm{FSO}}}(\bar{C}^{\mathrm{RF}})$ and $\mathbb{E}_g\{C^{\mathrm{FSO}}|C^{\mathrm{FSO}}\leq \bar{C}^\mathrm{RF}\}$, which are determined in the following. The CDF of the achievable rate  of the FSO channel is a function of the CDF of  a squared  Hoyt distributed RV, $u^2$, \cite{Hoyt}  and  is given by 
\begin{IEEEeqnarray}{lll}\label{CDF_FSO}
F_{C^{\mathrm{FSO}}}(x)&=1-\frac{2m}{1+m^2}\, \textrm{Ie}\left(\!\frac{1-m^2}{1+m^2},\frac{(1+m^2)^2}{4m^2\Omega}\chi(x)\!\right),\quad\chi(x)\geq0,\qquad
\end{IEEEeqnarray}
where $\chi(x)=\frac{-k_g w^2}{4}\ln\left(\frac{1}{\bar{\gamma}^2}\left({e^{\frac{2\ln(2)x}{W^\mathrm{FSO}}}-1}\right)\right),  0\leq x\leq\frac{W^\mathrm{FSO}}{2}\log_2(\bar{\gamma}+1)$, and the Rice-$\textrm{Ie}$ function is defined as
\begin{IEEEeqnarray}{rll}\label{CDF_Qfunction}
\textrm{Ie}(v,t)=\frac {1}{\sqrt {1-v^{2}}} \left [{Q(\sqrt {v_1t}, \sqrt {v_2t})-Q(\sqrt {v_2t}, \sqrt {v_1t})}\right], \qquad
\end{IEEEeqnarray}
where $v_1=1+\sqrt{1-v^2}$, $v_2=1-\sqrt{1-v^2}$, and $Q(\cdot, \cdot)$ denotes the Marcum $Q$-function.
For the special case, where the {FSO} beam is orthogonal to the PD plane, i.e., $m=1$, (\ref{CDF_FSO})  becomes the CDF of an exponential distribution. This result is in line with  \cite{ICC}, where for an orthogonal beam,  misalignment factor $u$ was shown to be Rayleigh distributed, which implies that $u^2$ is exponentially distributed.
\begin{theo}\label{Theorem3}
For low and high SNRs, the average {FSO}  rate, conditioned on   the FSO channel being the instantaneous bottleneck of   the end-to-end achievable rate,  is given by 
\iftoggle{OneColumn}{\begin{IEEEeqnarray}{rll}\label{Theo3}
\mathbb{E}^\mathrm{low}_g\{C^{\mathrm{FSO}}|C^{\mathrm{FSO}}\leq \bar{C}^\mathrm{RF}\} 
		= \frac{W^\mathrm{FSO}(1+m^2)}{4\pi\ln(2) m\Omega } \sum_{\ell=1}^\infty\frac{(-\bar{\gamma}^2)^{\ell}}{\ell} e^{-a\chi(\bar{C}^\mathrm{RF})}\int\limits_0^\pi\frac{e^{b\chi(\bar{C}^\mathrm{RF}) \cos t}}{a-b\cos t} \mathrm{d}t,\quad \bar{\gamma}<1,\qquad\IEEEyesnumber\IEEEyessubnumber\label{theo3}
\end{IEEEeqnarray}}
{\begin{IEEEeqnarray}{rll}\label{Theo3}
\mathbb{E}&^\mathrm{low}_g\{C^{\mathrm{FSO}}|C^{\mathrm{FSO}}\leq \bar{C}^\mathrm{RF}\} 
= \frac{W^\mathrm{FSO}(1+m^2)}{4\pi\ln(2) m\Omega } \nonumber\\&\times\sum_{\ell=1}^\infty\frac{(-\bar{\gamma}^2)^{\ell}}{\ell} e^{-a\chi(\bar{C}^\mathrm{RF})}\int\limits_0^\pi\frac{e^{b\chi(\bar{C}^\mathrm{RF}) \cos t}}{a-b\cos t} \mathrm{d}t,\quad \bar{\gamma}<1,\qquad\IEEEyesnumber\IEEEyessubnumber\label{theo3}
\end{IEEEeqnarray}}
\iftoggle{OneColumn}{\begin{IEEEeqnarray}{rll}
		\mathbb{E}^\mathrm{high}_g\{C^{\mathrm{FSO}}|C^{\mathrm{FSO}}\leq \bar{C}^\mathrm{RF}\} 
		&=\frac{W^\mathrm{FSO}}{2\ln(2)}\Big(\ln\left(\bar{\gamma}^2\right)F_{C^{\mathrm{FSO}}}(\bar{C}^{\mathrm{RF}})-\frac{2(1+m^2)e^{-\delta\chi(\bar{C}^\mathrm{RF})}}{\pi k_gw^2m\Omega}\nonumber \\ & \times\int\limits_{0}^\pi   \frac{e^{b\chi(\bar{C}^\mathrm{RF})\cos t}(1+\chi(\bar{C}^\mathrm{RF})(\delta-b\cos t))}{(\delta-b\cos t)^2} \mathrm{d}t\Big),\quad\bar{\gamma}\gg1,\IEEEyessubnumber\label{theo3_I}
\end{IEEEeqnarray}}
{\begin{IEEEeqnarray}{rll}
	&\mathbb{E}^\mathrm{high}_g\{C^{\mathrm{FSO}}|C^{\mathrm{FSO}}\leq \bar{C}^\mathrm{RF}\} =\frac{W^\mathrm{FSO}}{2\ln(2)}
	\nonumber\\&\times\Big(\!\ln\left(\!\bar{\gamma}^2\!\right)F_{C^{\mathrm{FSO}}}(\bar{C}^{\mathrm{RF}})-\frac{2(1+m^2)e^{-\delta\chi(\bar{C}^\mathrm{RF})}}{\pi k_gw^2m\Omega} \nonumber\\&\times\int\limits_{0}^\pi\!\!   \frac{e^{b\chi(\bar{C}^\mathrm{RF})\cos t}(1+\chi(\!\bar{C}^\mathrm{RF}\!)(\delta-b\cos t))}{(\delta-b\cos t)^2} \mathrm{d}t\!\Big),\!\!\bar{\gamma}\!\gg\!1,\qquad\IEEEyessubnumber\label{theo3_I}
\end{IEEEeqnarray}}
where $a=\frac{4\ell}{k_gw^2}+\delta$, $b=\frac{(1-m^4)}{4m^2\Omega}$, and $\delta=\frac{(1+m^2)^2}{4m^2\Omega}$. 
 \end{theo}
\begin{IEEEproof}
The proof is given in Appendix \ref{App4}.
\end{IEEEproof}

In Section \ref{Sec_sim}, we will show that the infinite sum in (\ref{theo3})  converges to the exact result if only the first $5$ terms are used. Moreover, (\ref{theo3}) and (\ref{theo3_I}) include  finite-range integrals that can be calculated numerically. Furthermore, it can be shown that Theorem \ref{Theorem2} is a special case of Theorem \ref{Theorem3}. In particular, if the RF channel always supports a higher achievable rate than the FSO channel, or equivalently if the conditions in the  expected values in (\ref{theo3}) and (\ref{theo3_I}) are always fulfilled by assuming $\bar{C}^\mathrm{RF}\rightarrow\infty$, then (\ref{Theo3})  approaches the ergodic rate of the FSO channel in Theorem \ref{Theorem2}.

In summary, we can closely approximate the ergodic sum rate for UAVs employing non-BA relaying by substituting (\ref{theo1}), (\ref{CDF_FSO}),  (\ref{theo3}), and (\ref{theo3_I}) into (\ref{lem2}) to obtain 
\iftoggle{OneColumn}{\begin{IEEEeqnarray}{rll}\label{non-BA-result}
		\lim\limits_{K\rightarrow \infty}\bar{C}^\mathrm{NB}_\mathrm{low}&= \mathbb{E}^\mathrm{low}_g\{C^{\mathrm{FSO}}|C^{\mathrm{FSO}}\leq \bar{C}^\mathrm{RF}\}+\bar{C}^{\mathrm{RF}}(1-F_{C^{\mathrm{FSO}}}(\bar{C}^{\mathrm{RF}}))	, \quad  \bar{\gamma}< 1,\qquad\IEEEyesnumber\IEEEyessubnumber\label{non_BA_result_low}\\
		\lim\limits_{K\rightarrow \infty}\bar{C}^\mathrm{NB}_\mathrm{high}&= \mathbb{E}^\mathrm{high}_g\{C^{\mathrm{FSO}}|C^{\mathrm{FSO}}\leq \bar{C}^\mathrm{RF}\}+\bar{C}^{\mathrm{RF}}(1-F_{C^{\mathrm{FSO}}}(\bar{C}^{\mathrm{RF}}))	, \quad  \bar{\gamma}\gg1.\qquad\IEEEyessubnumber\label{non_BA_result_high}
\end{IEEEeqnarray}}
{\begin{IEEEeqnarray}{rll}\label{non-BA-result}
\lim\limits_{K\rightarrow \infty}\bar{C}^\mathrm{NB}_\mathrm{low}&=\mathbb{E}^\mathrm{low}_g\{C^{\mathrm{FSO}}|C^{\mathrm{FSO}}\leq \bar{C}^\mathrm{RF}\}\nonumber\\&+\bar{C}^{\mathrm{RF}}(1-F_{C^{\mathrm{FSO}}}(\bar{C}^{\mathrm{RF}})), \quad  \bar{\gamma}< 1,\IEEEyesnumber\IEEEyessubnumber\label{non_BA_result_low}\\
\lim\limits_{K\rightarrow \infty}\bar{C}^\mathrm{NB}_\mathrm{high}&= \mathbb{E}^\mathrm{high}_g\{C^{\mathrm{FSO}}|C^{\mathrm{FSO}}\leq \bar{C}^\mathrm{RF}\}\nonumber\\&+\bar{C}^{\mathrm{RF}}(1-F_{C^{\mathrm{FSO}}}(\bar{C}^{\mathrm{RF}})), \quad  \bar{\gamma}\gg1.\qquad\IEEEyessubnumber\label{non_BA_result_high}
\end{IEEEeqnarray}}
Eq.~(\ref{non-BA-result}) together with  (\ref{theo1}), (\ref{CDF_FSO}), and (\ref{Theo3}) reveals that the ergodic rate for  non-BA relaying at the UAV crucially depends on the position  and the instability of the UAV via $\bar{\gamma}$, ${\gamma}$,  $m$, and $\Omega$. In Section \ref{Sec_sim}, we investigate the accuracy of (\ref{non-BA-result}) for small numbers of MUs and employ this expression to study the performance of non-BA relaying   UAV-based communications networks and to optimize the position of the non-BA relaying UAV.


\section{ Simulation Results}\label{Sec_sim}

In this section, first we use simulation results  to validate the analytical results in (\ref{BA_result}) and (\ref{non-BA-result}) for   {BA} and  non-{BA} relay UAVs, respectively. Then, we investigate the impact of the positioning of the UAV on the system performance. Finally, we study the inherent trade-offs in the considered network and the impact of different system and channel parameters on the end-to-end ergodic sum rate. 

For  the considered system and channel models, the parameter values  provided in Table \ref{Table:Sys_Param} are adopted, unless specified otherwise.  The {MU}s are homogeneous Poisson distributed with density $\lambda=0.008$ $\mathrm{MUs}/\mathrm{m^2}$ over a circular area with radius $r_0=50\ $m, where the center of this area is located at $(x_{0}, y_{0}, 0)=(600,0,0)$ m. We allocate $W^\mathrm{RF}_\mathrm{sub}=\frac{W^\mathrm{RF}}{\bar{K}}=79.4\ \mathrm{kHz}$ to each MU where $W^\mathrm{RF}=5\ \mathrm{MHz}$ and $\bar{K}=\mathbb{E}\{K\}=\lambda \pi r_0^2=63.3$. The GS is located  at a height of $z_\mathrm{GS}=100$ m above the origin of the Cartesian coordinate system. 
\begin{table}[t]
	\caption{System and Channel Parameters \cite{Drone_coverage}, \cite{Hourani_opt_alt}, \cite{ICC}.}
	\centering
	\scalebox{0.7}{%
		\begin{tabular}{|| l | c | c || }
			\hline  
			{FSO Channel Parameters} &{Symbol}& {Value}  \\ 
			\hline
			\hline
			FSO bandwidth &$W^{\text{FSO}}$ &$1\ \mathrm{GHz}$\\
			Aperture radius &$a$&$10\ \mathrm{cm}$\\
			Beam waist radius &$w_0$& $0.25\ \mathrm{mm}$\\ 
			Responsivity &$R$ &$0.5$ \\
			Index of refraction structure parameter &$C^2_n(0)$& $1.7 \times 10^{-14}\ \mathrm{m}^{2/3}$\\
			Tx power &$\bar{p}$&$0.1\ \mathrm{mW}$\\
			Noise variance &$\rho^2$&$10^{-14}\ \mathrm{A}^2/\mathrm{Hz}$\\
			Attenuation factor&$\kappa$&$16.8\times 10^{-3} \frac{\mathrm{dB}}{m}$\\
			GS height &$z_\mathrm{GS}$ &$100\ \mathrm{m}$\\
			\hline
			\hline
			{RF Channel Parameters}&& \\
			\hline 
			\hline
			Operating frequency &$f$ &$2\ \mathrm{GHz}$\\
			RF bandwidth &$W^{\text{RF}}_\mathrm{sub}$&$79\ \mathrm{kHz}$\\
			User transmit power &$P$ &$20\ \mathrm{dBm}$\\
			Multipath power &$\eta^2$ &$22\ \mathrm{dBm}$\\
			Noise power spectral density &$N_0$&$-114\ \mathrm{dBm}/\mathrm{MHz}$\\
			Shadowing parameters &$(\mu,\sigma^2)$ &$(0,3)\ \mathrm{dB}$\\
			Radius of {MU}s' area	&$r_0$&$50\ \mathrm{m}$\\
			{LOS} probability  &$(B,C)$ &$(0.136,60.69)$\\
			MU density &$\lambda$ &$0.008$ $\mathrm{MUs}/\mathrm{m}^2$\\
			\hline
			\hline
			{UAV Parameters}&&  \\
			\hline
			\hline
			Position STD &$\sigma_{p}$&$1\ \mathrm{cm}$\\
			Orientation STD &$\sigma_{o}$&$0.3\ \mathrm{mrad}$\\
			Number of RF antennas		&$N$&$2$\\
			Initial Position &$(x_{0},y_{0},z_{d0})$&$(600,0,30)\ \mathrm{m}$\\
			\hline
		\end{tabular}
	}
	\label{Table:Sys_Param}
\end{table}

For the RF channel, the two state channel model in (\ref{EQ:RFchannel}) with LOS and NLOS states is adopted. In the NLOS and LOS states,  Rayleigh fading with multipath power $\eta^2$ and  shadowed Rician fading with lognormal shadowing parameters $(\mu,\sigma)$ are assumed, respectively. The  LOS probability parameters, $(B,C)$, in (\ref{LOS_prob}) are chosen for an urban  environment \cite{Hourani_opt_alt}.
For the FSO channel, simulations with and without GG  atmospheric turbulence were conducted. The atmospheric loss and GML are incorporated according to (\ref{FSO_Channel}). The  UAV is assumed to be able to track the PD with zero mean misalignment factor, $u$, and the UAV's instability in the hovering state is accounted for by the position and orientation standard deviations (STD) $\sigma_o=0.3\ \mathrm{mrad}$ and $\sigma_p=1\ \mathrm{cm}$, respectively.   
Given the above assumptions and parameters, we obtained the results reported in this section by averaging over $10^6$ realizations of the RF and FSO channels as well as of the MU distributions.

\subsection{Validation of  Analytical Results}
In this subsection, we investigate the accuracy of the following assumptions and approximations made in our analysis:
1) ignorance of the atmospheric turbulence, $g_a$, in the FSO channel, 2) accuracy of $\bar{C}^\mathrm{NB}_\mathrm{sum}$ in (\ref{lem2}) for finite $K$,
3) fitting of  the lognormal shadowing to Nakagami fading in (\ref{dist_tau}), 4) Gaussian-Chebyshev Quadrature (GCQ) numerical approximation in (\ref{theo1}), 5)  Taylor series expansions in (\ref{BA_result_low}) and (\ref{non_BA_result_low}). Furthermore, we confirm our analytical results in  (\ref{BA_result}) and (\ref{non-BA-result}) for  low and high SNR scenarios.

\begin{figure}[t]
	\centering
	\includegraphics[width=0.7\textwidth]{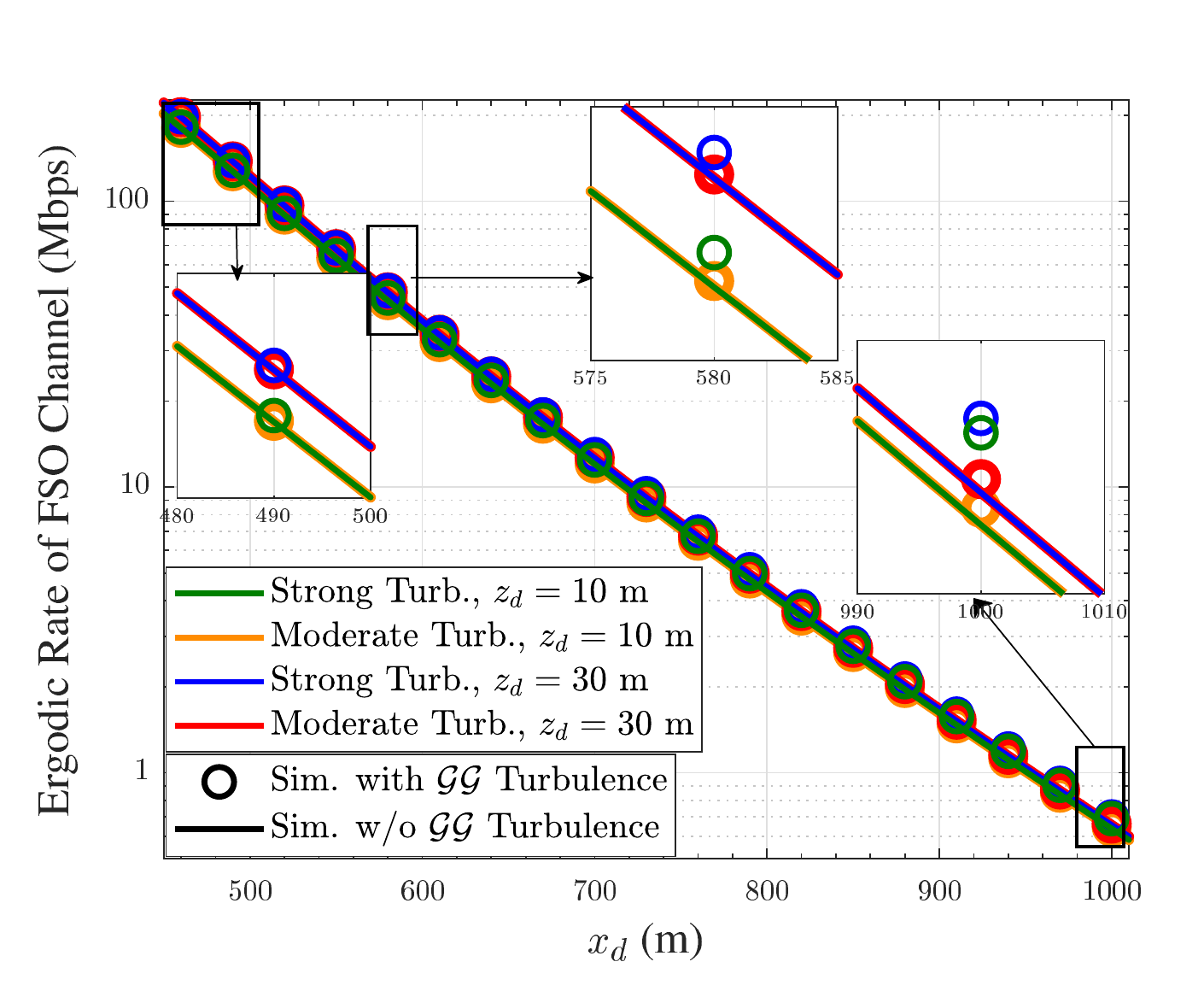}
	\vspace{-4mm}
	\caption{Ergodic rate of the FSO channel vs. distance of the UAV from the GS under strong ($C_n^2(0)=10^{-13}$) and moderate  ($C_n^2(0)=10^{-14}$) turbulence conditions and  UAV  altitudes of $z_d=10, 30$ m.} \vspace{-2mm}
	\label{Fig:GGapprox}
\end{figure}

In Section \ref{subsec_GGassum}, we argued that the atmospheric turbulence factor, $g_a$, can be ignored when the UAV flies at typical operating altitudes and for small distances from the GS. Fig.~\ref{Fig:GGapprox} investigates the accuracy of this approximation  by comparing  the ergodic rates of the FSO channel with and without GG fading as functions of the distance between the UAV and the GS.  Here, strong and moderate turbulence conditions\footnote{We note that the severity of the turbulence affects only the variance of the atmospheric turbulence factor $g_a$, while its mean is always one.  In fact, the normalized variance of $g_a$, i.e., the scintillation index $\left({\text{var}\{g_a\}\over  \mathbb{E}\{g_a\}^2}\right)$, varies for different  operating distances $L$ and different $C_n^2(z)$ \cite{Strong_Turb}, and depending on the operating scenario,  stronger turbulence can lead to larger/smaller variances  than  moderate turbulence. This means that, as can be observed in Fig.~\ref{Fig:GGapprox}, the ergodic rate of the FSO channel in strong turbulence is not always lower than that in moderate turbulence.}, which have different  ground level refraction structure parameters, $C_n^2(0)$, and different UAV operating altitudes,  $z_d$,  are considered. Fig.~\ref{Fig:GGapprox} suggests that at a distance of $1$ km, the gap between the curves with and without GG fading at  altitudes of $10$ m  and $30$ m is respectively about  $7 \%$ and $5 \%$ for strong turbulence. The smaller gap for the higher altitude is due to  the H-V model since $C_n^2(z_d)$ decreases if $z_d$ increases. Furthermore, Fig.~\ref{Fig:GGapprox} shows that the  gap between the ergodic rates with and without GG fading vanishes for small  distances. Considering the dependence of parameters $\alpha$ and $\beta$ on the distance in (\ref{alphabeta}), the impact of atmospheric turbulence decreases for shorter distances. Since the practical operating range of  UAVs   is expected to be within one kilometer of the GS and  its typical operating altitude is expected to be above $30$ m, the impact of atmospheric turbulence can be savely ignored.

\begin{figure}[t]
	\centering
	\includegraphics[width=0.7\textwidth]{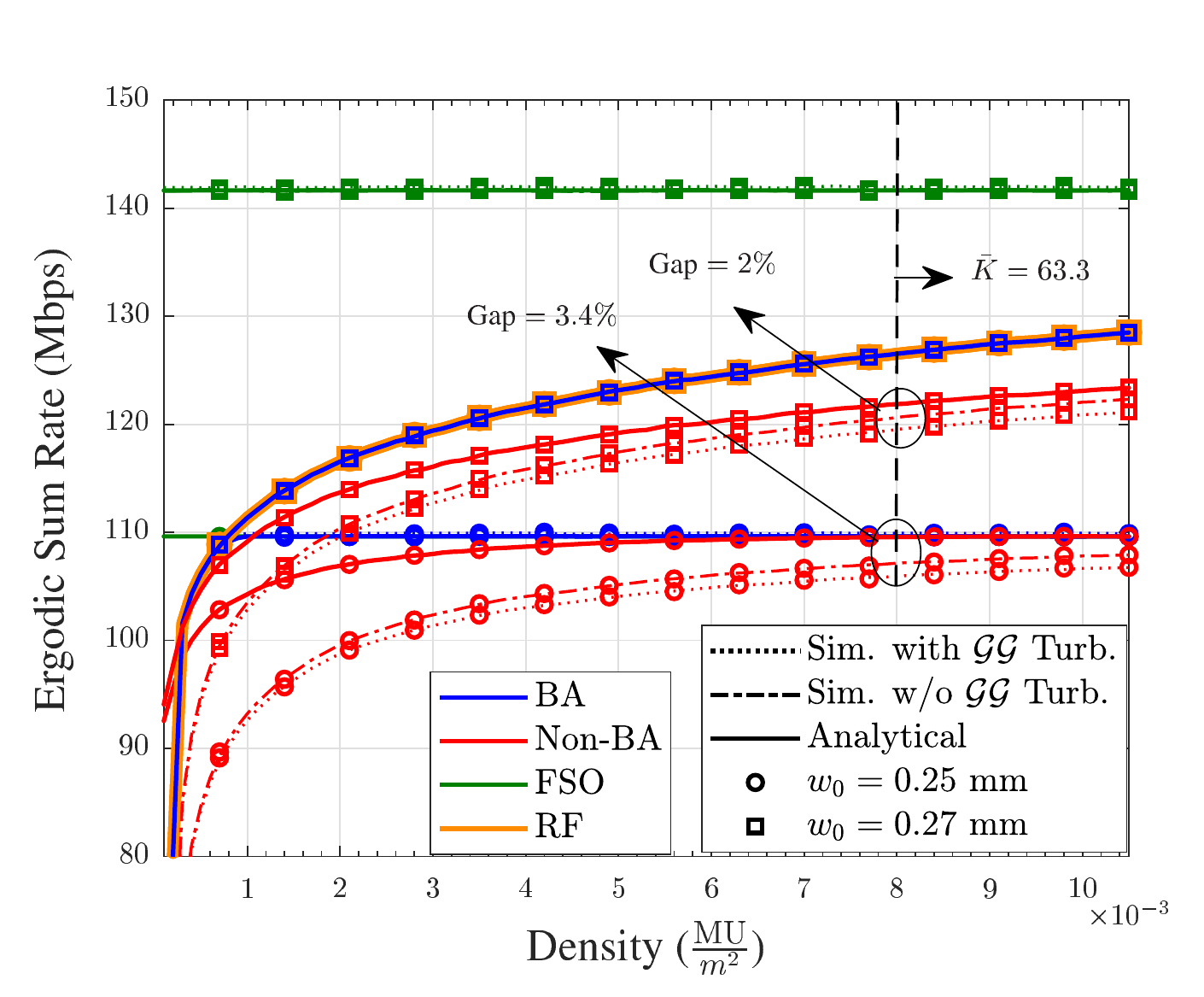}
	\vspace{-4mm}
	\caption{Ergodic sum rate vs.\ MU density for BA and non-BA relay UAVs for $W_\mathrm{RF}^\mathrm{sub}=\frac{W_\mathrm{RF}}{\bar{K}}$ and $\bar{K}=\lambda \pi r_0^2$. The analytical results  for the FSO channel, the RF channel,  BA relaying, and  non-BA relaying are obtained from (\ref{theo2}), (\ref{theo1}),   (\ref{non_BA_result_low}), and  (\ref{BA_result_low}), respectively.} \vspace{-2mm}
	\label{Fig:RatevsDens}
\end{figure}

Fig.~\ref{Fig:RatevsDens} shows the ergodic sum rate  versus  MU density, $\lambda$, for FSO  beam waists of $w_0=w({L=0})=0.25$ mm and $0.27$ mm. Here,  the subchannel bandwidth assigned to the MUs is proportional to the average number of MUs and  the total bandwidth is kept constant for all values of $\lambda$, i.e., $W^\mathrm{RF}_\mathrm{sub}=\frac{W^\mathrm{RF}}{\bar{K}}$ where $W^\mathrm{RF}=5 \mathrm{MHz}$ and $\bar{K}=\mathbb{E}\{K\}=\lambda \pi r_0^2$. Fig.~\ref{Fig:RatevsDens} confirms that not taking into account $g_a$ does not affect the ergodic rate of the FSO channel and yields accurate  results for  BA relaying. For  non-BA relaying, there is a small gap of about $1\%$ between the simulation results with and without GG fading. Furthermore, in the non-BA case, the simulation results are upper bounded by the analytical results obtained from (\ref{non-BA-result}). However, for sufficiently large numbers of  MUs or equivalently for sufficiently high MU densities, the instantaneous sum rate of the RF channel  approaches the ergodic sum rate of the RF channel (see Lemma \ref{Lemma2}) and hence, the gap between numerical and analytical results vanishes also for  non-BA relaying. Fig.~\ref{Fig:RatevsDens} suggests that even for a relatively small number of MUs, this gap is  small. For example, for $\lambda=0.008$ or equivalently  $\bar{K}=63.3\ \mathrm{MU}$, the gap is only about $2\%$ and $3.4\%$ for $w_0=0.25$ mm and $w_0=0.27$ mm, respectively. 


\begin{figure*}[t]
	\centering
	\begin{minipage}[b]{0.45\textwidth}
		\centering
		\hspace{-0.5cm}
		\includegraphics[width=1\textwidth]{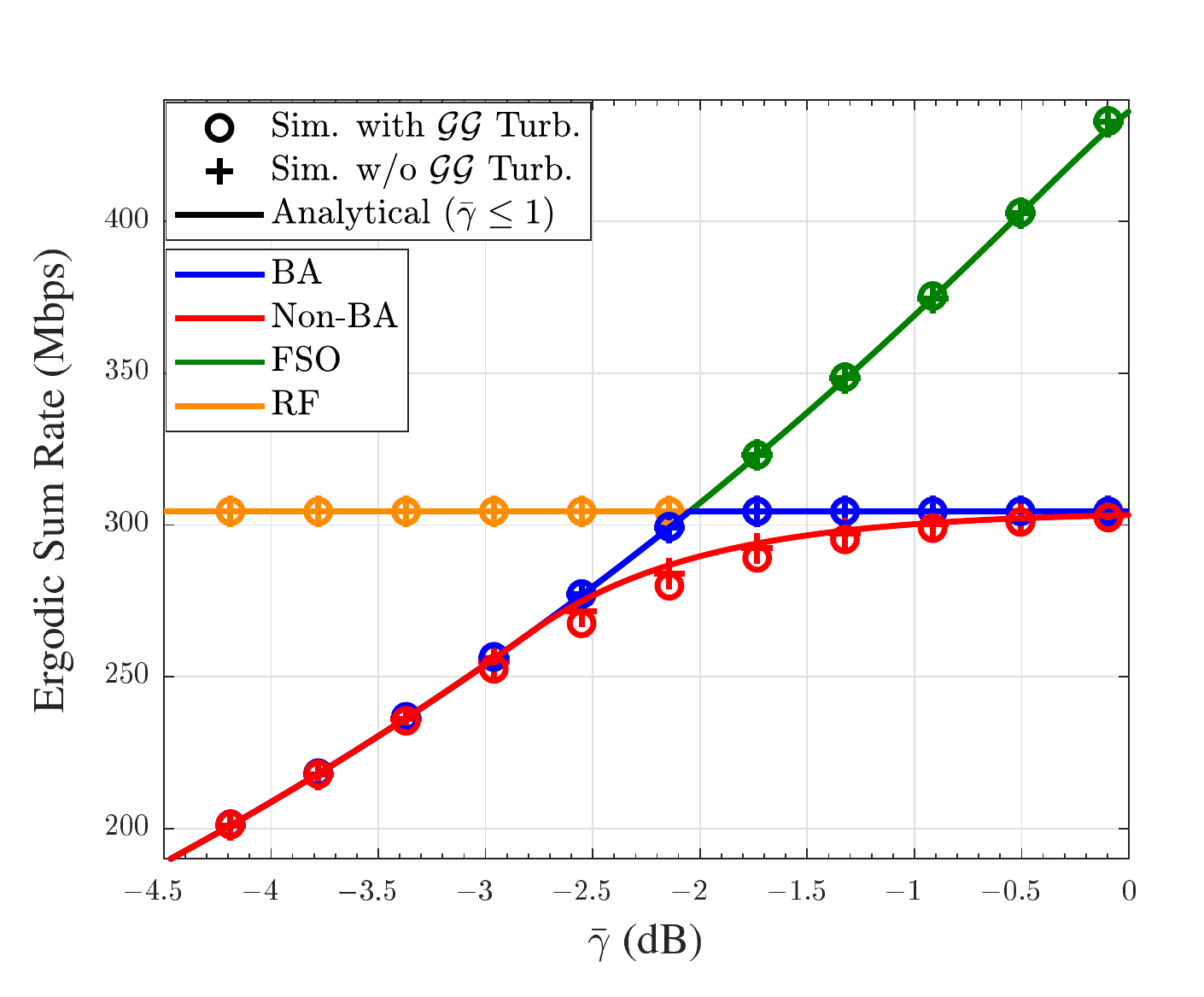}  \vspace{-9mm}
		\caption{Ergodic sum rate of RF and  FSO channels and  BA and non-BA relaying vs. SNR of FSO channel in the low SNR regime. Analytical results for FSO channel (\ref{theo2}), RF channel (\ref{theo1}), BA relaying (\ref{BA_result_low}), and non-BA relaying (\ref{non_BA_result_low}) are shown.  $r_0=80$ m.} 
		\label{Fig:Ratelowsnr}
	\end{minipage}
	\hfill
	\begin{minipage}[b]{0.01\textwidth}
	\end{minipage}
	\hfill
	\begin{minipage}[b]{0.45\textwidth}
		\centering
		\hspace{-0.5cm}
		\includegraphics[width=1\textwidth]{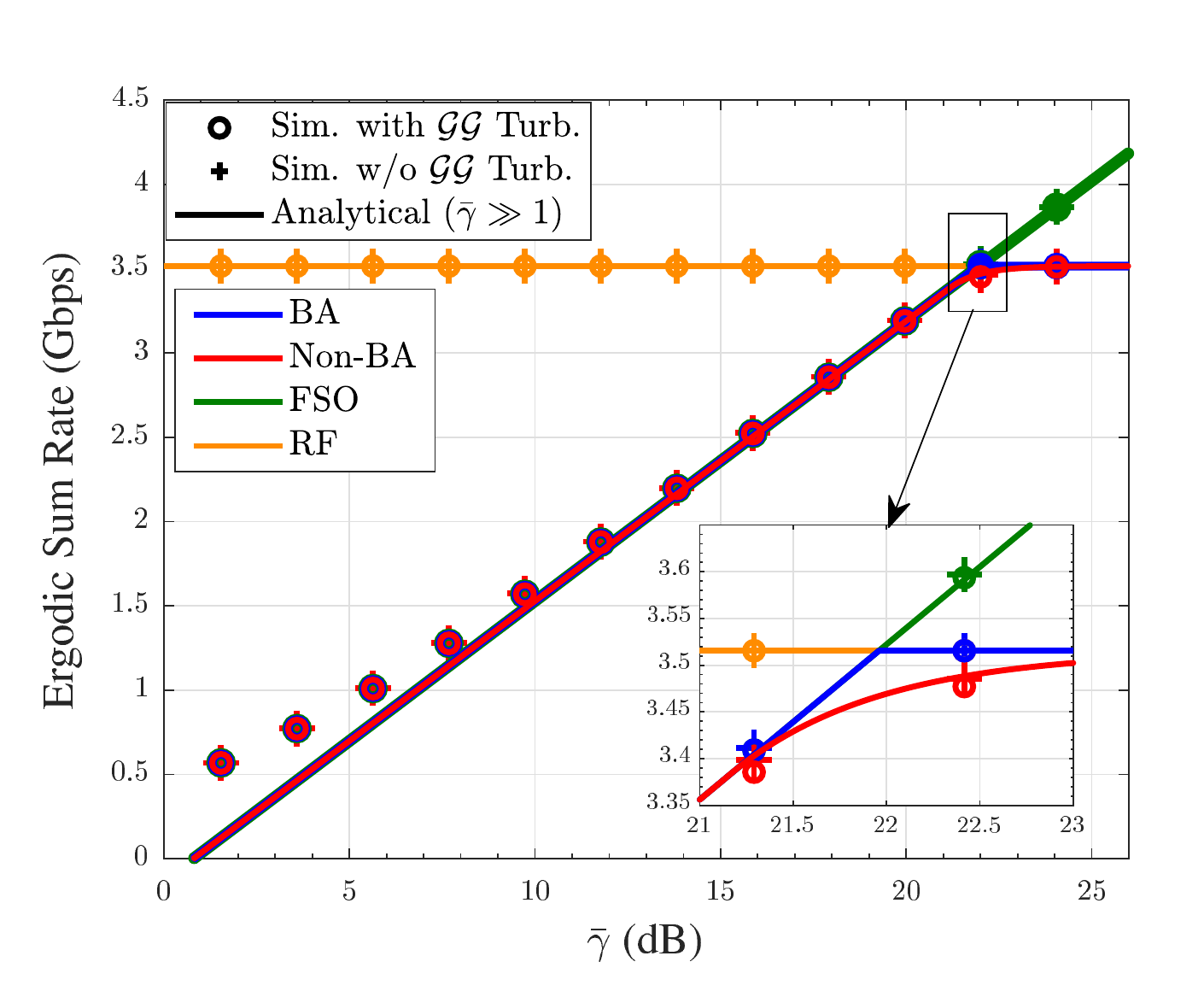}   \vspace{-9mm}
		\caption{Ergodic sum rate of RF and  FSO channels and  BA and non-BA relaying vs. SNR of FSO channel in the high SNR regime. Analytical results for FSO channel (\ref{theo2_II}), RF channel (\ref{theo1}), BA relaying (\ref{BA_result_high}), and non-BA relaying (\ref{non_BA_result_high}) are shown.  $r_0=300$ m.} 
		\label{Fig:Ratehighsnr}
	\end{minipage}
	\hfill
	\begin{minipage}[b]{0.00\textwidth}
	\end{minipage} \vspace{-5mm}
\end{figure*}
In Figs.~\ref{Fig:Ratelowsnr} and \ref{Fig:Ratehighsnr}, we compare the analytical and simulation results for low and high SNRs, respectively, and make the following observations.  1) Although we consider only the first five terms in the summation of the Taylor series in (\ref{theo2}), the analytical ergodic rate of the FSO channel perfectly matches the corresponding simulation results. 2) The analytical ergodic sum rate of the RF channel, where we  consider only 10 terms ($H=M=10$) for the GCQ approximation in (\ref{theo1}), agrees well with the simulation results. This also confirms that the approximation of the log-normal distribution by Nakagami fading for shadowed Rician fading in (\ref{dist_tau}) is justified. Overall, we conclude that the analytical ergodic  sum rate expressions for  BA and non-BA relaying UAVs in (\ref{BA_result_high})  and  (\ref{non_BA_result_high}), respectively, are accurate.


\subsection{Impact of Positioning of the UAV}

\begin{figure}[t!]
	\centering
	\includegraphics[width=0.7\textwidth,]{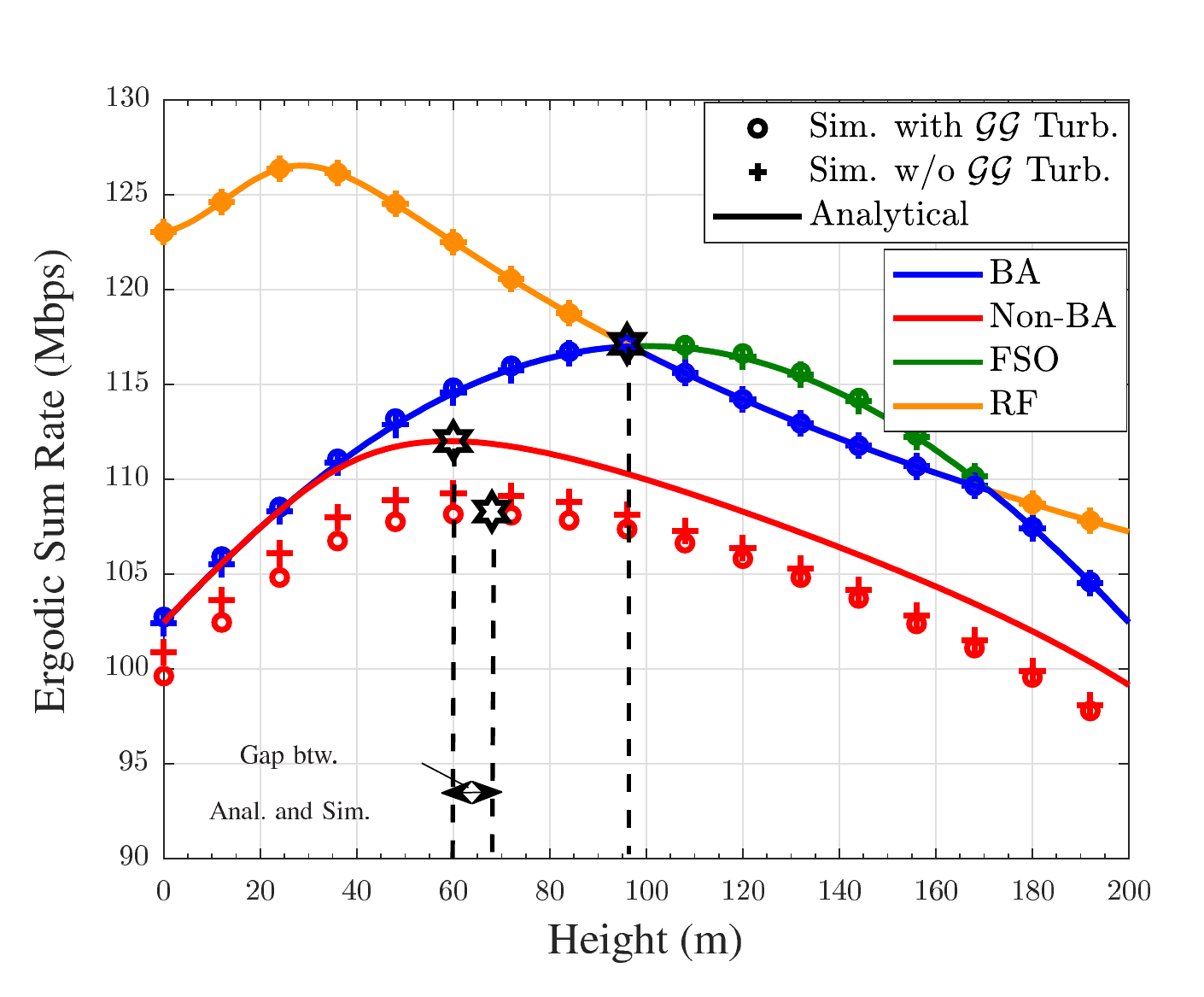}
	\vspace{-4mm}
	\caption{Ergodic sum rate vs.\ UAV altitude ($z_d$) when the UAV hovers above the center of the MUs' area. Analytical results for the FSO channel   (\ref{theo2}), the RF channel  (\ref{theo1}),  BA relaying  (\ref{BA_result_low}),  non-BA relaying  (\ref{non_BA_result_low}) are shown. } \vspace{-2mm}
	\label{Fig:Ratevsheight}
\end{figure}

Next, we investigate the impact of the placement of the UAV on the end-to-end ergodic rate. Fig.~\ref{Fig:Ratevsheight} depicts the ergodic sum rate as a function of the UAV's altitude, where the UAV is located at the center of the MUs' area. 
Here, the ergodic sum rate of the {RF} link  suggests an optimum altitude of $30\ \mathrm{m}$, which is a direct consequence of the trade-off between the probability of a LOS link and the value of the path loss. Furthermore, the ergodic rate of the FSO channel reaches its maximum value at a height of $100$ m; the same height at which the {PD} is installed. At this altitude, the FSO beam is orthogonal w.r.t. the {PD} plane and accordingly, the GML, $g_g$, and the atmospheric loss, $g_p$, assume their respective minimum values, cf. Remark \ref{Remark 4}. The optimum height for {BA} relaying, $z_d^\mathrm{BA}$, depends on  the altitude  at which the ergodic sum rate curves of both links intersect, i.e., $\bar{C}^\mathrm{RF}=\bar{C}^\mathrm{FSO}$ and is marked by $\davidsstar$ in Fig.~\ref{Fig:Ratevsheight}. On the other hand, for  non-BA relaying, the gap between the  analytical and simulation results  is only $2\%$ and the altitudes that maximize the respective curves, $z_d^\mathrm{NB}$, are only 8 m apart. Because of the benefits of buffering, BA relaying yields a  higher ergodic sum rate than  non-BA relaying.

In Fig. \ref{Fig:RatevsPos},  the UAV  operates at a height of  $30$ m and moves along the $x$-axis from the cell center towards the GS. By reducing the distance to the GS, the ergodic rate of the FSO channel drastically increases, due to the exponential reduction of the atmospheric loss. On the other hand, the ergodic rate of the RF channel decreases  due to the larger path loss and the smaller LOS probability caused by the smaller elevation angle in (\ref{LOS_prob}). Consequently, farther from the cell center,  the RF channel is the performance  bottleneck and the ergodic rates of both {BA} and  non-BA relaying approach the ergodic sum rate of the RF channel. On the other hand, when the UAV is farther from the {GS}, the {FSO} channel limits the performance of both types of relaying.
For BA relaying, the intersection of the RF and FSO ergodic sum rate curves corresponds to the optimal position of the UAV at $11$ m, and analysis and simulations yield the same value. On the other hand, comparing the analytical and simulated ergodic rates for non-BA relaying reveals a gap of only $1\%$. The maxima of the corresponding curves are 5 m apart, which has little impact on the optimal value since the ergodic rate curves are flat around the maxima.  Figs.~\ref{Fig:Ratevsheight} and  \ref{Fig:RatevsPos} confirm that the optimal positioning of the relay depends on the parameters of the RF and FSO channels as well as the type of relaying.
\iftoggle{OneColumn}{\begin{figure}[t!]
		\centering
		\includegraphics[width=0.7\textwidth]{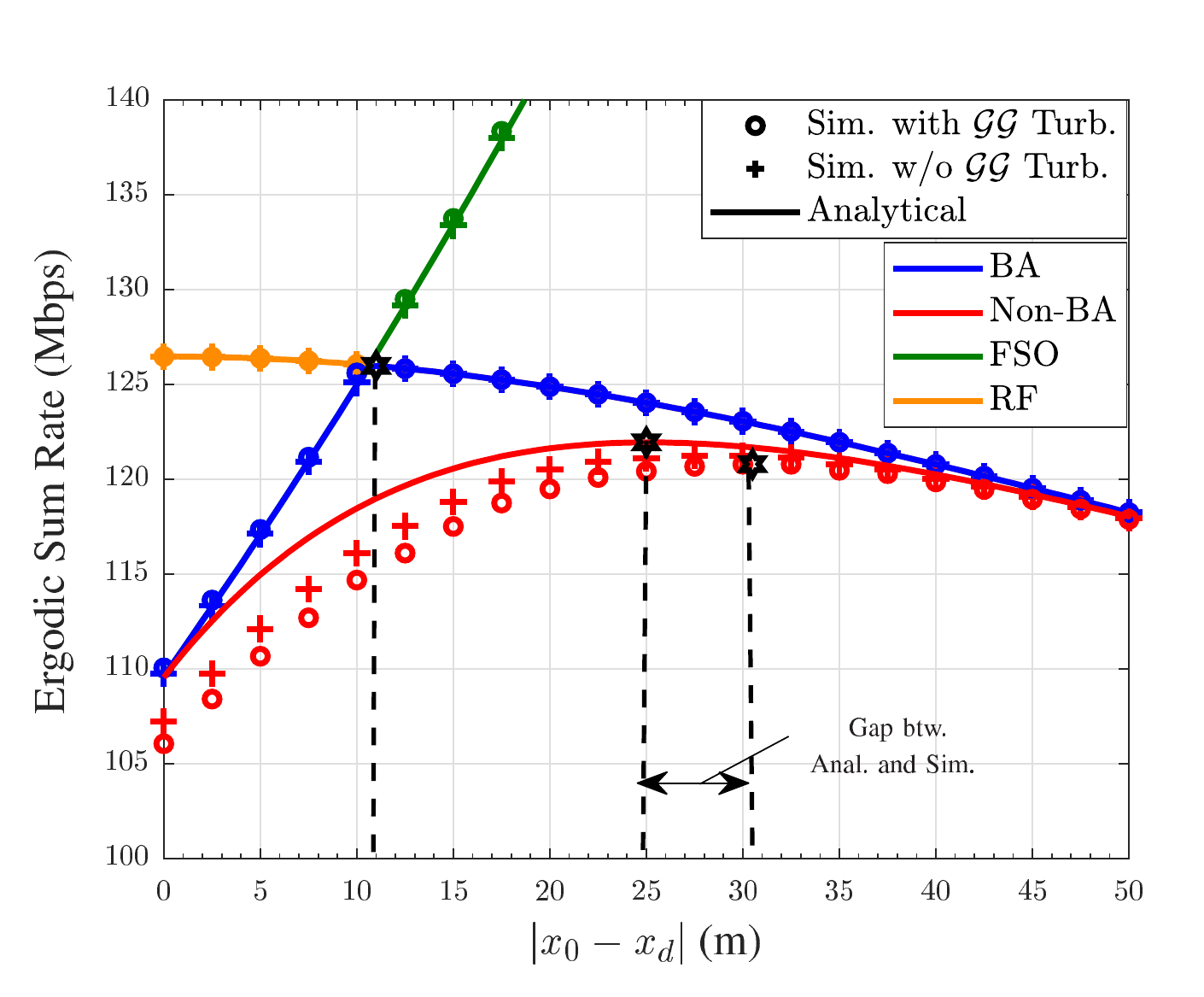} 
		\vspace{-4mm}
		\caption{Ergodic sum rate vs.\ UAV's distance from the center of the MUs' cell. UAV moves along x-axis. Analytical results  for the FSO channel   (\ref{theo2}), the RF channel  (\ref{theo1}),  BA relaying  (\ref{BA_result_low}),  non-BA relaying (\ref{non_BA_result_low}) are shown.} \vspace{-2mm}
		\label{Fig:RatevsPos}
\end{figure}}{\begin{figure}[t]
		\centering
		\resizebox{0.9\linewidth}{!}{\psfragfig{Fig/Fig3/RatePos}{
				\psfrag{OPT}[c][c][0.9]{ }
				\psfrag{Gap}[c][c][0.9]{Gap btw.}
				\psfrag{GapI}[c][c][0.88]{$x_d^\mathrm{BA} \text{and} \ x_d^\mathrm{NB}$}
				\psfrag{GapII}[c][c][0.9]{Gap  btw.}
				\psfrag{GapIII}[c][c][0.9]{Anal. and Sim.}
		}} 
		\vspace{-4mm}
		\caption{Ergodic sum rate vs.\ UAV's distance from the center of the MUs' cell. UAV moves along x-axis. Analytical results  for the FSO channel   (\ref{theo2}), the RF channel  (\ref{theo1}),  BA relaying  (\ref{BA_result_low}),  non-BA relaying (\ref{non_BA_result_low}) are shown.} \vspace{-2mm}
		\label{Fig:RatevsPos}
	\end{figure}
}

\subsection{Impact of System Parameters}
\iftoggle{OneColumn}{\begin{figure}[t]
		\centering
		\includegraphics[width=0.7\textwidth]{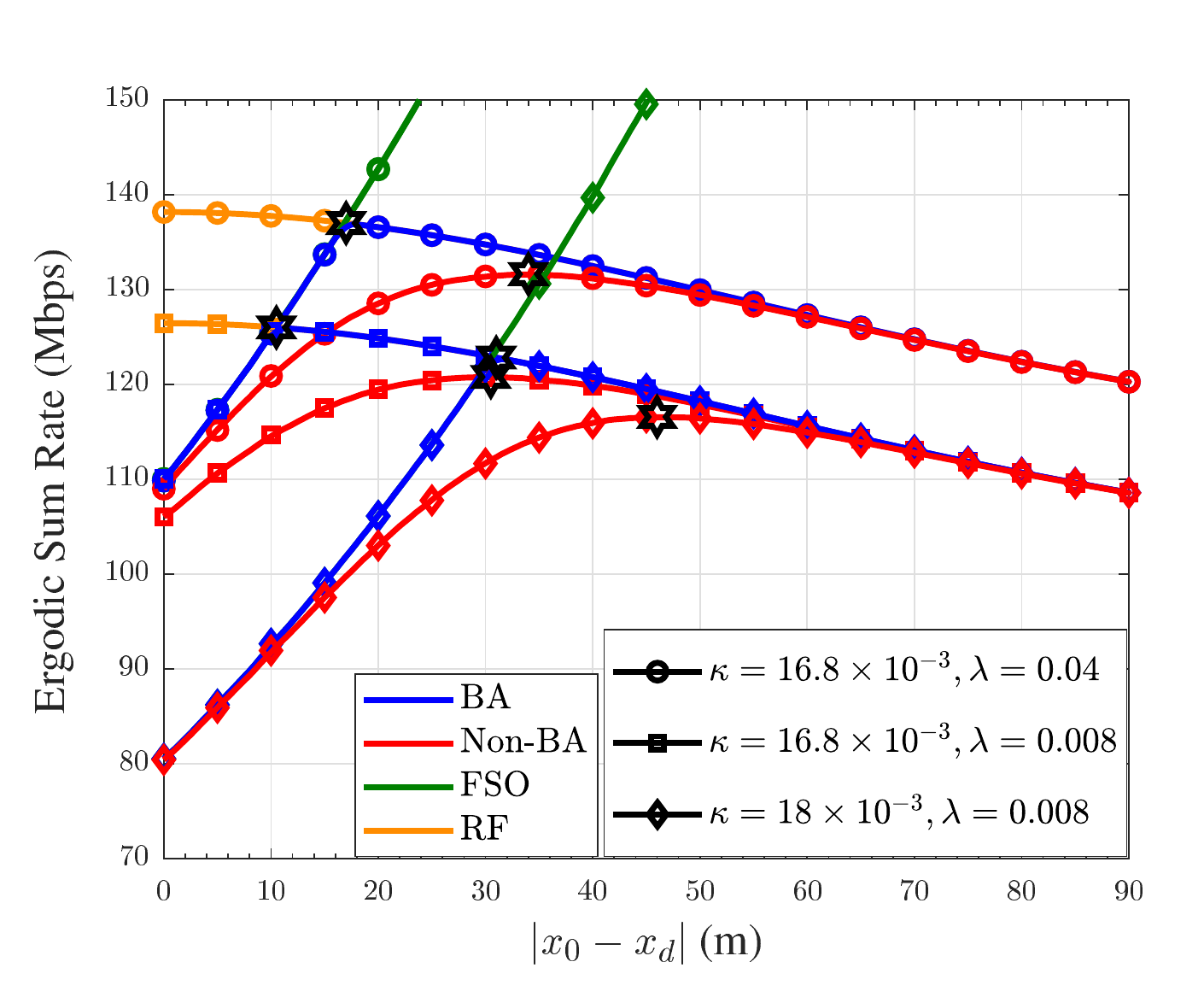}
		\vspace{-4mm}
		\caption{Simulated ergodic sum rate vs.\ UAV's distance from cell center for attenuation factors $\kappa=[16.8, 18]\times 10^{-3} \frac{\mathrm{dB}}{m}$ and MU densities $\lambda=[0.008, 0.04]\ \mathrm{MUs}/\mathrm{m}^2$.} \vspace{-2mm}
		\label{Fig:RatevsPos_kap}
\end{figure}}{
	\begin{figure}
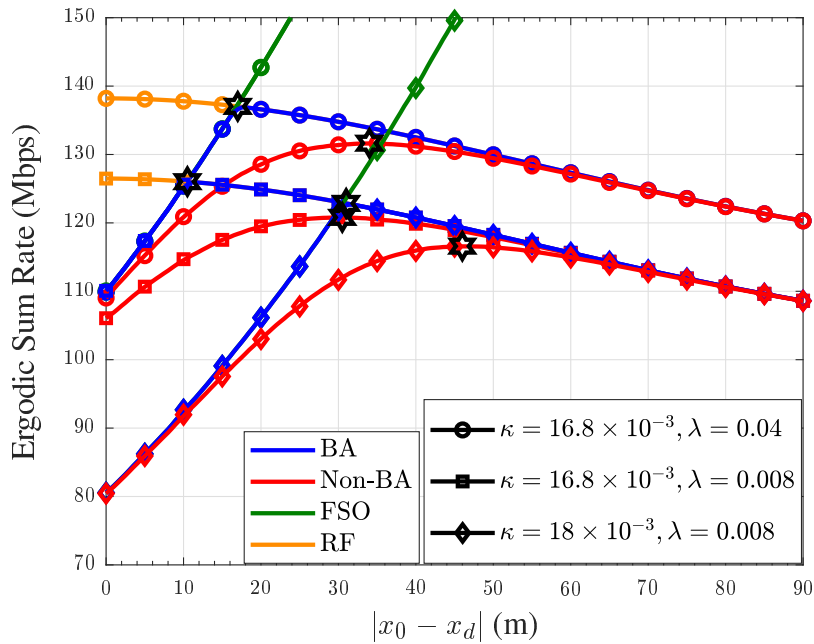

		\centering
		\resizebox{0.9\linewidth}{!}{\psfragfig{Fig/Fig5/RatePos4}{
				\psfrag{Kappa}[c][c][1.2]{$\kappa\uparrow$}
				\psfrag{Lambda}[c][c][1.2]{$\lambda \uparrow$}
		}} 
		\vspace{-4mm}
		\caption{Simulated ergodic sum rate vs.\ UAV's distance from cell center for attenuation factors $\kappa=[16.8, 18]\times 10^{-3} \frac{\mathrm{dB}}{m}$ and MU densities $\lambda=[0.008, 0.04]\ \mathrm{MUs}/\mathrm{m}^2$.} \vspace{-2mm}
		\label{Fig:RatevsPos_kap}
	\end{figure}
}

In this subsection, we investigate the impact of   the weather-dependent FSO attenuation factor, $\kappa$, and the density of the MUs, $\lambda$, on the end-to-end system performance. Here, for clarity of presentation, we only show simulated ergodic sum rates.

Fig.~\ref{Fig:RatevsPos_kap} shows the impact of  different  weather-dependent FSO attenuation factors, i.e.,  $\kappa=[16.8, 18]\times 10^{-3} \frac{\mathrm{dB}}{m} $, and different MU densities, i.e., $\lambda=[0.008,0.04]\ {\mathrm{MU}}/{\mathrm{m}^2}$, on the system performance and the optimum position of the UAV. As expected, for larger attenuation factors, the ergodic rate of the FSO link degrades due to the larger atmospheric loss.  Thus,  for larger $\kappa$, for both BA and non-BA relay UAVs, a position closer to the GS (i.e., larger $|x_0-x_d|$) is preferable  in order to compensate for the reduced ergodic rate in the FSO link.
 Furthermore, for higher densities $\lambda$, the bandwidth available  for each MU decreases, and accordingly, the $\mathrm{SNR}=\frac{P}{N_0 W_\mathrm{sub}^\mathrm{RF}}$ for each MU increases. Thus, the ergodic rate of  the RF channel improves. Although, a larger number of MUs does not affect the ergodic rate of the FSO channel,  the  FSO backhaul has to support the increased rate of the RF channel. Therefore, for larger MU densities, for both BA and non-BA relaying the UAV benefits from moving towards the GS (i.e., increasing $|x_0-x_d|$) to improve the quality of its backhaul channel. 

\section{Conclusions}\label{Sec_concl}
In this paper, the end-to-end performance of   a UAV-based communication system, where a relay UAV connects MUs via RF access links to a GS via an FSO backhaul link, was analyzed in terms of the ergodic sum rate. The UAV's characteristics including its relative position w.r.t. the GS and the MUs and its (in)stability in the hovering state  were taken into account in the adopted RF and FSO channel models. In particular, the probability of attaining an LOS path in the RF channel and the GML in the FSO channel were  accounted for. Furthermore, to address the application dependent sensitivity to delay, both BA and non-BA relay UAVs were investigated. Exact and approximate expressions for the ergodic sum rate  were derived for {BA} and  non-{BA} relay UAVs, respectively.  We  validated the accuracy of the obtained analytical result and investigated the trade-offs affecting the optimum position of  BA and non-BA relay UAVs. Our results revealed that the impact of atmospheric turbulence on the quality of the FSO channel can be ignored for practical UAV-GS distances of less than 1 km. Furthermore, the derived approximate analytical expression for the ergodic sum rate for non-BA relays was shown to  approach the corresponding simulation results  for  sufficiently large numbers of MUs. Moreover, our simulations revealed that  the random variations of  the FSO channel  caused by the UAV's instability can be mitigated by BA relaying which results in larger achievable ergodic sum rate compared to non-BA relaying at the expense of introducing an additional delay into the system. Our results also show that when the weather conditions get worse and accordingly the atmospheric loss in the backhaul channel increases, the UAV prefers a  position closer to the GS  to improve the quality of the backhaul channel. Furthermore, for higher  MU densities and  the resulting  larger  amounts of  data received via the RF access channel, the end-to-end performance can be improved if the UAV moves closer to the GS in order to enhance  the backhaul link quality. Considering our simulation and analytical results, we conclude that the specific properties of both the FSO backhaul and  RF access channels have to be  simultaneously  taken into account for performance evaluation and optimization of   UAV-based communication networks employing mixed RF-FSO channels.

\appendices
\renewcommand{\thesectiondis}[2]{\Alph{section}:}

\section{Proof of Lemma~\ref{Lemma3}}\label{App5}
Given that $h_{k,n}^{sr}=h_k^s+h^{r}_{k,n} e^{j\Psi_{k,n}}$ is a Rician distributed RV with  Nakagami distributed parameter $h_k^s\sim\mathrm{Nakagami}(q,\omega)$, the conditional distribution of $\sqrt{\tau}$  is given by a non-central chi distribution as follows
\begin{IEEEeqnarray}{rll}\label{ERgodic_shadow}
	&f_{\sqrt{\tau}|h_k^s}\left(x\right)= \frac{x^N}{\eta^{2}\varkappa^{N-1}}\exp\left(-\frac{x^2+\varkappa^2}{2\eta^2}\right)I_{N-1}\left(\frac{\varkappa x}{\eta^2}\right),\qquad
\end{IEEEeqnarray}
where $\varkappa^2=\sum\limits_{n=1}^N (h^s_k)^2=N (h^s_k)^2$. Since $h^s_k\sim\mathrm{Nakagami}\left(q,\omega\right)$, thus, $(h^s_k)^2$ is a  Gamma distributed RV and the pdf of  $\varkappa^2$ is given by    
\begin{IEEEeqnarray}{rll}
	&f_{\varkappa^2}\left(y\right)= \frac{q^q}{\left(\omega N\right)^q\Gamma (q)} y^{q-1}e^{-qy/\left(N\omega\right)}.\quad
\end{IEEEeqnarray}
Then, we can obtain the unconditional  distribution of $\sqrt{\tau}$ as $f_{\sqrt{\tau}}\left(x\right)=\int\limits_{0}^{\infty}f_{\sqrt{\tau}|h_k^s}\left(x\right)f_{\varkappa^2}\left(y\right) \mathrm{d}y $.  To solve this integral, we exploit  $I_v(x)=\frac{(x/2)^v}{v!}\,_0\mathcal{F}_1\left(v+1,x^2/4\right)$  \cite{Sum_of_Squared} and \cite[Eq.~(7.522-9)]{integral}, where $_0\mathcal{F}_1\left(\cdot,\cdot\right)$ is the confluent hypergeometric function. Thus, the pdf of $\sqrt{\tau}$  is given by
\begin{IEEEeqnarray}{rll}
	&f_{\sqrt{\tau}}\left(x\right)= \frac{2\left(\frac{N\omega}{2q\eta^2}+1\right)^{-q} x^{2N-1}}{(2\eta^2)^{N}\left(N-1\right)!e^{\frac{x^2}{2\eta^2}}} \,_1\mathcal{F}_1\left(q,N;\frac{x^2}{2\eta^2+\frac{4\eta^4 q}{N\omega}}\right).\nonumber\\
\end{IEEEeqnarray}
Then, using the relation $f_\tau(x)=\frac{1}{2\sqrt{x}}f_{\sqrt{\tau}}(\sqrt{x})$,  the pdf of $\tau$ is obtained as in (\ref{dist_tau}) and this concludes the proof.

\section{Proof of Theorem~\ref{Theorem1}}\label{App1}
First, the ergodic rate corresponding to (\ref{inst_RF_rate}) can be simplified  to a summation of the ergodic rates for the LOS and NLOS states as follows 
\begin{IEEEeqnarray}{rll}\label{Lemma1}
	\bar{C}^\mathrm{RF}&=\mathbb{E}_{h,\Phi}\{C^{\mathrm{RF}}\}=W_{\mathrm{sub}}^{\mathrm{RF}} \ \mathbb{E}_{h,\Phi}\left\{\sum\limits_{(r_k,\phi_k)\in \Phi} R^{\mathrm{RF}}_k\right\}\nonumber\\
	&\overset{(a)}{=}W_{\mathrm{sub}}^{\mathrm{RF}}\, \mathbb{E}_{\Phi}\left\{\!\sum\limits_{(r_k,\phi_k)\in \Phi}\mathbb{E}_{h_{\mathrm{r}}}\{\!R_k^{\mathrm{RF}}\!\}\mathrm{P}_{\!\mathrm{NLOS},k}\!+\!\mathbb{E}_{h_{\mathrm{sr}}}\{R_k^{\mathrm{RF}}\}\mathrm{P}_{\mathrm{LOS},k}\!\right\},
	\qquad
\end{IEEEeqnarray}
where for equality $(a)$, we exploited the linearity of expectation.

Then, using Campbell's law,   which states that $\mathbb{E}\left(\sum\limits_{x\in\Phi} f(x)\right)=\lambda \int\limits_{\mathcal{R}^2} f(x) \mathrm{d}x$, where $\Phi$ is a homogeneous Poisson  process   with intensity $\lambda$ and $f$ is  any nonnegative function \cite{Haenggi}, we obtain
\begin{IEEEeqnarray}{rll}\label{Proof_theo1_I}
\mathbb{E}_{\Phi,h}\{\!C^{\mathrm{RF}}\!\}&=\lambda W_{\mathrm{sub}}^{\mathrm{RF}}\int\limits_0^{2\pi}\int\limits_{0}^{r_0} \Big(\mathbb{E}_{h_{\mathrm{r}}}\{R_k^{\mathrm{RF}}\}\mathrm{P}_{\mathrm{NLOS},k}\nonumber\\ &+\mathbb{E}_{h_{\mathrm{sr}}}\{R_k^{\mathrm{RF}}\}\mathrm{P}_{\mathrm{LOS},k}\Big) r_k \mathrm{d}r_k \mathrm{d}\varphi_k.
\end{IEEEeqnarray}
The above integrals can  be solved only numerically. To obtain a suitable numerical approximation, we first change variables $r_k$ and $\phi_k$ to $x=\frac{2r_k}{r_0}-1$ and $y=\frac{\phi_k}{\pi}-1$, respectively, which yields 
\begin{IEEEeqnarray}{rll}\label{Proof_theo1_II}
	\mathbb{E}_{\Phi,h}\{C^{\mathrm{RF}}\}&=\lambda W_{\mathrm{sub}}^{\mathrm{RF}}\ \int\limits_{-1}^{1}\int\limits_{-1}^{1} \Big(\mathbb{E}_{h_{\mathrm{r}}}\{R_k^{\mathrm{RF}}\}\mathrm{P}_{\mathrm{NLOS},k}\nonumber\\ &+\mathbb{E}_{h_{\mathrm{sr}}}\{R_k^{\mathrm{RF}}\}\mathrm{P}_{\mathrm{LOS},k}\Big) x \mathrm{d}x \mathrm{d}y.
\end{IEEEeqnarray}
Then, we note that any definite integral can be transformed  to a weighted summation using Gaussian-Chebyshev Quadrature (GCQ) as  $\int\limits_{-1}^{+1} f(x)\frac{1}{\sqrt{1-x^2}}\mathrm{d}x=\varpi \sum\limits_{i=1}^{H} f(x_i)+\tilde{E}_H$, where  $\varpi=\frac{\pi}{H}$ and $ x_i=\cos\left(\frac{2i-1}{2H}\pi\right)$ \cite{stegun}. Thus, letting $f(x,y)= x \sqrt{(1-x^2)}\times \linebreak[0]\sqrt{(1-y^2)} \big(\mathbb{E}_{h_{\mathrm{r}}} \{R_k^{\mathrm{RF}}\} \mathrm{P}_{\mathrm{NLOS},k}$ $ +\mathbb{E}_{h_{\mathrm{sr}}}\{R_k^{\mathrm{RF}}\}\mathrm{P}_{\mathrm{LOS},k}\big)$,  (\ref{Proof_theo1_II}) can be written as
\begin{IEEEeqnarray}{rll}\label{Proof_theo1_III}
	&\mathbb{E}_{\Phi,h}\{C^{\mathrm{RF}}\}=\lambda W_{\mathrm{sub}}^{\mathrm{RF}}\ \int\limits_{-1}^{1}\int\limits_{-1}^{1} f(x,y) \frac{1}{\sqrt{1-x^2}}\frac{1}{\sqrt{1-y^2}} \mathrm{d}x \mathrm{d}y\nonumber\\
	&\overset{(a)}{=}\lambda W_{\mathrm{sub}}^{\mathrm{RF}}\ \int\limits_{-1}^{1} \left(\frac{\pi}{H}\sum\limits_{i=1}^{H} f(x_i,y)+\tilde{E}_H\right) \frac{1}{\sqrt{1-y^2}}  \mathrm{d}y\nonumber\\
	&\overset{(b)}{=}\lambda W_{\mathrm{sub}}^{\mathrm{RF}}\left(\frac{\pi}{H} \sum\limits_{i=1}^{H} \int\limits_{-1}^{1}  f(x_i,y)\frac{1}{\sqrt{1-y^2}}\mathrm{d}y\right)+{E}_H  \mathrm\nonumber\\
	&=\frac{\pi^2\lambda W_{\mathrm{sub}}^{\mathrm{RF}}}{HM} \sum\limits_{j=1}^{M}\sum\limits_{i=1}^{H} f(x_i,y_j)+E_H+E_M,
\end{IEEEeqnarray}
where in $(a)$ and $(b)$ the GCQ is applied to the integrals over $x$ and $y$, respectively.
  
Next, we determine the ergodic rate for  shadowed Rician fading, denoted by $\mathbb{E}_{h_{\mathrm{sr}}} \{R_k^{\mathrm{RF}}\}=\mathbb{E}_{h_{\mathrm{sr}}} \left\{\log_2(1+\gamma\lVert{\mathbf{\tilde{h}}}_k\rVert^2)\right\}$ as follows
  \begin{IEEEeqnarray}{rll}\label{Gamma}
  	&\mathbb{E}_{h_{\mathrm{sr}}} \{R_k^{\mathrm{RF}}\}=\int\limits_0^\infty \log_2(1+\gamma x) f_{{\tau}}\left(x\right) \mathrm{d}x,
  \end{IEEEeqnarray}
 where $f_{{\tau}}\left(x\right)$ is given in Lemma \ref{Lemma3}.  The  above integral can be solved using the identity $\,_1\mathcal{F}_1\left(a,b;x\right)=\sum\limits_{n=0}^\infty\frac{(a)_n}{n!(b)_n}x^n$ \cite{Rayleigh_rate} and \cite[Eq.~(78)]{Sum_of_Squared}. This leads to   the ergodic sum rate for  shadowed Rician fading in (\ref{Rate_sr}).
 Then, using \cite[Eq.~(40)]{Rayleigh_rate} for the ergodic sum rate for Rayleigh fading leads to (\ref{Rate_r,sr}) and this concludes the proof.
    
 \section{Proof of Theorem~\ref{Theorem2}}\label{App2}
 Based on (\ref{FSO_cap_inst}), the ergodic rate of the FSO channel is given by
\begin{IEEEeqnarray}{rll}\label{Proof_theo2_I}
&\mathbb{E}_g\{ C^\mathrm{FSO}\}=\frac{W^\mathrm{FSO}}{2}\mathbb{E}_g\left\{\log_2\left(1+\frac{e\bar{p}^2}{2\pi\rho^2} g^2\right)\right\}.
\end{IEEEeqnarray}
By substituting $g=R_s  g_p g_g$  and $g_g$ from (\ref{g_g}), we obtain
\begin{IEEEeqnarray}{rll}\label{Proof_theo2_II}
\mathbb{E}_g\!\{\!C^\mathrm{FSO}\!\}\!=\!\frac{W^\mathrm{FSO}}{2}\mathbb{E}_{u^2}\left\{\!\log_2\left(\!1+ \bar{\gamma}^2  \exp\left(\!-\frac{4u^2}{k_gw^2}\!\right)\!\right)\!\right\}.\qquad
\end{IEEEeqnarray}
Then, for low SNRs, we use the Taylor series $\ln(1+x)=\sum\limits_{\ell=1}^\infty\frac{(-1)^\ell}{\ell}x^\ell$ (for $|x|\leq 1$) to obtain 
\begin{IEEEeqnarray}{rll}\label{Proof_theo2_III}
&\bar{C}_\mathrm{low}^\mathrm{FSO}=\frac{W^\mathrm{FSO}}{2\ln(2)}\sum_{\ell=1}^{\infty}\frac{(-1)^{\ell}}{\ell}(\bar{\gamma})^{2\ell} \mathbb{E}_{u^2}\left\{\exp\left(-\frac{4\ell u^2}{k_gw^2}\right)\right\}.\qquad
\end{IEEEeqnarray}
The expectation over the Hoyt-squared variable, $u^2$, can be solved by \cite[6.611-4]{integral}. This leads to (\ref{theo2}).

For high SNRs, we use $\ln(1+x)\big|_{x\gg 1}{\approx} \ln(x)$ to obtain
\begin{IEEEeqnarray}{rll}\label{Proof_theo2_IV}
&\bar{C}_\mathrm{high}^\mathrm{FSO}=\frac{W^\mathrm{FSO}}{2}\left(\log_2\left(\bar{\gamma}^2\right)-\frac{4\mathbb{E}\{u^2\}}{\ln(2)k_gw^2}\right).
\end{IEEEeqnarray}

Substituting $\mathbb{E}_{u}\{u^2\}=\Omega$ into (\ref{Proof_theo2_IV}), we obtain (\ref{theo2_II}) which concludes the proof.

\section{Proof of Lemma~\ref{Lemma2}}\label{App3}
First, exploiting Jensen's inequality for concave functions, i.e., $\mathbb{E}\{f(x)\}\leq f(\mathbb{E}\{x\})$, the mutual independence of the FSO channel and  the RF  channels, and the MUs' random positions, we obtain
\begin{IEEEeqnarray}{rll}\label{proof_lem2_I}
\bar{C}_{\mathrm{sum}}^\mathrm{NB}&=\mathbb{E}_{h,\Phi,g}\{\min\left(C^{\mathrm{RF}},C^{\mathrm{FSO}}\right)\}\nonumber\\&\leq\mathbb{E}_{g}\{\min\big(\mathbb{E}_{h,\Phi}\{C^{\mathrm{RF}}\},C^{\mathrm{FSO}}\big)\},\end{IEEEeqnarray}
where  equality holds if $C^{\mathrm{RF}}=\mathbb{E}_{h,\Phi}\{C^{\mathrm{RF}}\}$. For $K\rightarrow\infty$, the instantaneous rate  in (\ref{inst_RF_rate}) and the ergodic rate of the RF channel in (\ref{theo1})  become identical since
  \begin{IEEEeqnarray}{rll}\label{RF_constant}
   \lim\limits_{K\rightarrow \infty}	C^{\mathrm{RF}}&= \lim\limits_{K\rightarrow \infty} W^{\mathrm{RF}}_{\mathrm{sub}}  \sum\limits_{(r_k,\phi_k)\in \Phi}R_k^{\mathrm{RF}}&\overset{(a)}{=}\mathbb{E}_{h,\Phi}\{C^{\mathrm{RF}}\},\quad
    \end{IEEEeqnarray}
  where  $(a)$ exploits the definition of the ergodic rate.  Given this relation, equality holds in (\ref{proof_lem2_I}).
Denoting $\mathbb{E}_{h,\Phi}\{C^{\mathrm{RF}}\}$ by $\bar{C}^{\mathrm{RF}}$, we obtain,
\begin{IEEEeqnarray}{rll}\label{proof_lem2_II}
\lim\limits_{K\rightarrow \infty}	\bar{C}_{\mathrm{sum}}^\mathrm{NB}&=\mathbb{E}_{g}\{\min\big(\bar{C}^{\mathrm{RF}},C^{\mathrm{FSO}}\big)\}\nonumber\\
	&\overset{(a)}{=}\!\mathbb{E}_{g}\{\bar{C}^{\mathrm{RF}}\mathrm{Pr}\left(C^{\mathrm{FSO}}\geq\bar{C}^{\mathrm{RF}}\right)+C^{\mathrm{FSO}}\mathrm{Pr}\left( C^{\mathrm{FSO}}\leq\bar{C}^{\mathrm{RF}}\right)\}\nonumber\\ 
&\overset{(b)}{=}\!\Big[\bar{C}^{\mathrm{RF}}\left(1-F_{C^\mathrm{FSO}}\left(\bar{C}^{\mathrm{RF}}\right)\right)+ \mathbb{E}_g\left\{C^{\mathrm{FSO}}\vert C^{\mathrm{FSO}}\leq\bar{C}^{\mathrm{RF}}\right\}\Big],\qquad
\end{IEEEeqnarray}
where in $(a)$ we apply $\min(\bar{c},x)= x \text{Pr}(x\leq \bar{c})+\bar{c} \, \text{Pr}(\bar{c}<x)$, where $\bar{c}$ is a constant.  In $(b)$, we substitute the definition of the CDF $ F_{C^\mathrm{FSO}}(x)=1-\mathrm{Pr}(C^{\mathrm{FSO}}\geq x)$. This concludes the proof.

\section{Proof of Theorem~\ref{Theorem3}}\label{App4}
For low SNRs,  we can use a conditional version of (\ref{Proof_theo2_III}) as follows
\begin{IEEEeqnarray}{rll}\label{proof_theo3_I}
&\mathbb{E}^\mathrm{low}_{g}\{ C^\mathrm{FSO}|C^{\mathrm{FSO}}\leq \bar{C}^\mathrm{RF}\} =\frac{W^\mathrm{FSO}}{2\ln(2)}\sum_{\ell=1}^{\infty}\frac{(-1)^{\ell}}{\ell}(\bar{\gamma})^{2\ell}\mathbb{E}_{u^2}\left\{\exp\left(-\frac{4\ell u^2}{k_gw^2}\right)\big\vert u^2\geq \chi(\bar{C}^\mathrm{RF})\right\}\nonumber\\
&\overset{(a)}{=} \frac{W^\mathrm{FSO}(1+m^2)}{2\ln(2) m \Omega }\sum_{\ell=1}^{\infty}\frac{(-1)^{\ell}}{\ell} \left(\bar{\gamma}\right)^{2\ell}\int\limits_{\chi(\bar{C}^\mathrm{RF})}^{\infty} e^{-ax} {I}_0(bx) \mathrm{d}x,\qquad
\end{IEEEeqnarray}
where in $(a)$, the pdf of a squared Hoyt RV, i.e., $f_{u^2}(x)=\frac{1+m^2}{2m\Omega}\exp\left(-\frac{(1+m^2)^2x}{4m^2\Omega}\right)I_0\left(\frac{(1-m^4)x}{4m^2\Omega}\right) $, is substituted. Next, we change the integration variable to $y=x-\chi(\bar{C}^\mathrm{RF})$, and  use the integral form  of the modified Bessel function, $I_0(x)=\frac{1}{\pi}\int_0^\pi e^{x \cos(t)} \mathrm{d}t$ \cite[8.431-5]{integral} to obtain
\begin{IEEEeqnarray}{rll}\label{proof_theo3_II}
\mathbb{E}^\mathrm{low}_g\{C^{\mathrm{FSO}}|C^{\mathrm{FSO}}\leq \bar{C}^\mathrm{RF}\} &=\frac{W^\mathrm{FSO}(1+m^2)}{2\pi\ln(2) m\Omega }\sum_{\ell=1}^{\infty}\frac{(-1)^{\ell}}{\ell}\left(\bar{\gamma}\right)^{2\ell} e^{-a\chi(\bar{C}^\mathrm{RF})}\nonumber\\&\times\int_{0}^\pi  e^{b\chi(\bar{C}^\mathrm{RF})\cos t}\int\limits_0^\infty e^{-(a-b\cos t) y}\mathrm{d}y \mathrm{d}t.
\end{IEEEeqnarray}
The inner integral  yields $\frac{1}{a-b\cos(t)}$ \cite[3.310]{integral}, where $a-b\cos(t)>0$,  which directly  leads to (\ref{theo3}).

For high SNRs,  we use $ \ln(1+x)\big|_{x\gg 1}{\approx} \ln(x)$ to obtain
\begin{IEEEeqnarray}{rll}\label{Proof_theo3_IV}
	 &\mathbb{E}^\mathrm{high}_{g}\{ C^\mathrm{FSO}|C^{\mathrm{FSO}}\leq \bar{C}^\mathrm{RF}\}  =\frac{W^\mathrm{FSO}}{2\ln(2)}\int\limits_{\chi(\bar{C}^\mathrm{RF})}^\infty\left(\ln\left(\bar{\gamma}^2\right)-\frac{4x}{\ln(2)k_gw^2}\right)f_{u^2}(x)\mathrm{d}x\nonumber\\
	  &=\frac{W^\mathrm{FSO}}{2\ln(2)}\times\Bigg(\ln\left(\bar{\gamma}^2\right)F_{C^{\mathrm{FSO}}}(\bar{C}^{\mathrm{RF}})-\frac{2(1+m^2)}{\pi k_gw^2 m \Omega}
	  \int\limits_{\chi(\bar{C}^\mathrm{RF})}^{\infty} x e^{-\delta x} {I}_0(bx) \mathrm{d}x\Bigg).\qquad
\end{IEEEeqnarray}
Then, substituting $y=x-\chi(\bar{C}^\mathrm{RF})$ and applying the integral form of the modified Bessel function,  we obtain
\begin{IEEEeqnarray}{rll}\label{proof_theo3_V}
	&\mathbb{E}^\mathrm{high}_g\{C^{\mathrm{FSO}}|C^{\mathrm{FSO}}\leq \bar{C}^\mathrm{RF}\} \nonumber\\ 
	&=\frac{W^\mathrm{FSO}}{2\ln(2)}\Big(\ln\left(\bar{\gamma}^2\right)F_{C^{\mathrm{FSO}}}(\bar{C}^{\mathrm{RF}})-\frac{2(1+m^2)}{\pi k_gw^2 m \Omega} e^{-\delta\chi(\bar{C}^\mathrm{RF})}\nonumber\\
	&\times \int\limits_{0}^\pi  e^{b\chi(\bar{C}^\mathrm{RF})\cos t}\int\limits_0^\infty \left(y+\chi(\bar{C}^\mathrm{RF})\right) e^{-(\delta-b\cos t) y}\mathrm{d}y \mathrm{d}t\Big),\qquad	
\end{IEEEeqnarray}
and  applying \cite[3.326-2, 3.310]{integral} to the inner integral leads to (\ref{theo3_I}).

\bibliographystyle{IEEEtran}
\bibliography{My_Citation_11-09-2019}
\end{document}